\documentclass[apj]{emulateapj}
\usepackage{apjfonts}

\def\gax{\gtrsim}

\shorttitle{New radio minihalos in cool-core clusters}
\shortauthors{S.~Giacintucci et al.}

\begin{document}

\setlength{\pdfpageheight}{\paperheight}
\setlength{\pdfpagewidth}{\paperwidth}


\title{New detections of radio minihalos in cool cores of galaxy clusters}

\author{Simona Giacintucci\altaffilmark{1,2}, 
Maxim Markevitch\altaffilmark{3,2},  
Tiziana Venturi \altaffilmark{4},
Tracy E. Clarke\altaffilmark{5}, 
Rossella Cassano\altaffilmark{4},
Pasquale Mazzotta \altaffilmark{6,7}}
\altaffiltext{1}{Department of Astronomy, University of Maryland,
  College Park, MD 20742, USA; simona@astro.umd.edu}
\altaffiltext{2}{Joint Space-Science Institute, University of Maryland, College Park,
MD, 20742-2421, USA}
\altaffiltext{3}{Astrophysics Science Division, NASA/Goddard Space
  Flight Center, Greenbelt, MD 20771, USA}
\altaffiltext{4}{INAF - Istituto di Radioastronomia, via Gobetti 101, I-40129 Bologna,
Italy}
\altaffiltext{5}{Naval Research Laboratory, 4555 Overlook Avenue SW, Code 7213, Washington, DC 20375, USA}
\altaffiltext{6}{Dipartimento di Fisica, Universit\'a di Roma Tor
  Vergata, Via della Ricerca Scientifica 1, I-00133, Rome, Italy}
\altaffiltext{7}{Harvard-Smithsonian Center for Astrophysics, 60 Garden Street, Cambridge, MA 02138}

\submitted{ApJ in press}


\begin{abstract}
Cool cores of some galaxy clusters exhibit faint radio ``minihalos.'' Their
origin is unclear; their study has been limited by their small number. We
undertook a systematic search for minihalos in a large sample of X-ray
luminous clusters with high-quality radio data. In this paper, we report
four new minihalos (A\,478, ZwCl\,3146, RXJ\,1532.9+3021 and A\,2204), and
five candidates, found in the reanalyzed archival {\em Very Large Array}\/
observations. The radio luminosities of our minihalos and candidates are in
the range $10^{23-25}$ W Hz$^{-1}$ at 1.4 GHz, consistent with this type of
radio sources. Their sizes ($40\div160$ kpc in radius) are somewhat smaller than
those of the previously known minihalos. We combine our new detections with
previously known minihalos, obtaining a total sample of 21 objects, and
briefly compare the cluster radio properties to the average X-ray
temperature and the total masses estimated from {\em Planck}. We find that
nearly all clusters hosting minihalos are hot and massive. Beyond that,
there is no clear correlation between the minihalo radio power and cluster
temperature or mass (in contrast with the {\em giant}\/ radio halos found in
cluster mergers, whose radio luminosity correlates with the cluster
mass). {\em Chandra}\/ X-ray images indicate gas sloshing in the cool cores
of most of our clusters, with minihalos contained within the sloshing
regions in many of them. This supports the hypothesis that radio-emitting
electrons are reaccelerated by sloshing. Advection
of relativistic electrons by the sloshing gas may also play a role in the
formation of the less-extended minihalos.

\end{abstract}


%

\section{Introduction}\label{sec:intro}

A number of relaxed, cool-core clusters of galaxies are known to
host centrally-located, diffuse synchrotron radio emission in their 
cores, that typically fills the cooling region ($r\sim 50-300$ kpc). 
These extended radio sources -- called minihalos -- have very low surface
brightness and possess steep radio spectra, $\alpha >1$ (for $S_{\nu} \propto 
\nu^{-\alpha}$). Their emission usually encompasses the radio source 
associated with the cluster dominant galaxy, but extends to much greater radii.
Only about ten clusters have confirmed minihalos \citep[and references 
therein]{2012A&ARv..20...54F}. These include the prototype minihalo in 
the Perseus cluster \citep[Sijbring 1993]{1992ApJ...388L..49B,1990MNRAS.246..477P},
whose emission occupies a large volume of the cool core, much larger than the inner 
$r\sim 30$ kpc region occupied by the well-known X-ray cavities filled by 
the lobes of 3C\,84 \citep[e.g.,][]{2011MNRAS.418.2154F}.

The origin of radio minihalos is unclear. Even though minihalos have active 
galaxies in their centers, the time needed by the electrons to diffuse from 
the galaxy across the cooling region ($\sim10^9$ years) is much longer than 
the radiative cooling timescale ($\sim 10^{7}-10^8$ years) for $\gamma\sim 10^4$ 
electrons that emit at the observed radio frequencies in the strong magnetic 
fields of the cool cores \citep[e.g.,][]{2002MNRAS.334..769T}. Thus, 
{\em in situ} production or reacceleration of relativistic electrons is 
required to explain minihalos. One possibility is inelastic collisions between 
relativistic cosmic ray protons and thermal protons, which can provide a continous 
injection of relativistic electrons 
\citep[``secondary'' or ``hadronic'' models,][]{2004A&A...413...17P,2007ApJ...663L..61F, 2010ApJ...722..737K, 2010arXiv1011.0729K, 2013MNRAS.428..599F}. 
Alternatively, reacceleration models \citep{2002A&A...386..456G,2004A&A...417....1G}
posit that the radio synchrotron emission arises from pre-existing, cooled 
relativistic electrons (for instance, injected by past activity of the central 
radio galaxy and/or hadronic collisions) that are reaccelerated to ultra-relativistic 
energies by turbulence in the intracluster medium (ICM). 

A key question for the latter scenario is the origin of the turbulence responsible 
for accelerating the electrons. An interesting possibility is that it is induced 
by sloshing gas motions, detected using high-resolution X-ray observations in 
the cool cores of most relaxed clusters. The observational signature of these motions 
is the spiral- or arc-shaped cold fronts \citep[e.g.,][]{2001ApJ...562L.153M,2001ApJ...555..205M,2003ApJ...596..190M,2007PhR...443....1M,2009ApJ...704.1349O,2010A&A...516A..32G,2010A&A...511A..15L}. These are believed to be contact discontinuities at the 
bounderies of the dense cool core that is sloshing in the deep potential well 
of the cluster in response to, e.g., a gravitational perturbation caused by
the infall of a small subcluster \citep{2006ApJ...650..102A}. Numerical simulations 
have shown that sloshing can generate significant turbulence in the core 
\citep{2004JKAS...37..571F,2012A&A...544A.103V,2013ApJ...762...78Z}.

The existence of a possible connection between gas sloshing and radio
minihalos was first hinted at by the discovery of a spatial correlation 
of the minihalo with X-ray cold fronts in two clusters \citep{2008ApJ...675L...9M}, 
where the diffuse radio emission appears confined to the region bounded by the 
fronts. A similar correlation is also observed in the Perseus cluster 
(Markevitch et al. in preparation). \cite{2013ApJ...762...78Z} used simulations 
to test the possibility of accelerating relativistic electrons via sloshing-driven
turbulence. Though theoretical uncertainties are large, they found that 
turbulence can be strong enough to reaccelerate low-$\gamma$ electrons to higher 
energies ($\gamma\sim10^4$) and produce diffuse radio emission with morphology, 
radio power and spectral index consistent with actual minihalos.

However, despite theoretical effort, our understanding of the origin and physical
properties of minihalos and their relation to the cluster core dynamics is still 
limited by the small number of known objects. We undertook a search for new minihalos 
in a large sample of galaxy clusters that combines radio observations from the 
{\em Very Large Array} ({\em VLA}) archive and {\em Giant Metrewave  Radio Telescope} 
({\em GMRT}; proprietary and archival) along with X-ray data from the {\em Chandra} 
archive. In this paper, we present new minihalos (and several candidates requiring 
better data for confirmation) found during re-analysis of the {\em VLA} observations. 
Statistical analysis of the cluster sample and detailed radio/X-ray study of 
specific interesting cases will be presented in forthcoming papers.

We adopt $\Lambda$CDM cosmology with H$_0$=70 km s$^{-1}$ Mpc$^{-1}$, 
$\Omega_m=0.3$ and $\Omega_{\Lambda}=0.7$.


\begin{table}[t]
\caption[]{List of clusters}
\begin{center}
\begin{tabular}{lrccc}
\hline\noalign{\smallskip}
\hline\noalign{\smallskip}
Cluster  name&  RA$_{\rm J2000}$ & DEC$_{\rm J2000}$ & $z$ & scale \\ 
    &  (h, m, s) &  ($^{\circ}, ^{\prime}, ^{\prime \prime}$) & & 
(kpc$/^{\prime \prime}$)  \\
\noalign{\smallskip}
\hline\noalign{\smallskip}
MACS\,J0159.8-0849\dotfill & 01 59 48.0 & $-$08 49 00 & 0.405 &  5.413 \\
MACS\,J0329.6-0211\dotfill & 03 29 40.8 & $-$02 11 54 & 0.450 & 5.759 \\ 
A\,478\dotfill             & 04 13 20.7 & +10 28 35   &  0.088 &  1.646 \\
ZwCl\,3146\dotfill         & 10 23 39.6 & +04 11 10   & 0.290 & 4.350 \\
A\,1795\dotfill            & 13 49 00.5 & +26 35 07   &  0.062  & 1.195 \\
RX\,J1532.9+3021\dotfill   & 15 32 54.4 & +30 21 11   & \phantom{0}0.362$^a$ & 5.048 \\
A\,2204\dotfill            & 16 32 45.7 & +05 34 43   &  0.152  &  2.643 \\
ZwCl\,1742.1+3306\dotfill  & 17 44 13.5 & +32 58 55   & 0.076 &  1.441 \\
MACS\,J1931.8-2635\dotfill & 19 31 48.0 & $-26$ 35 00 & 0.352  &  4.958 \\
\hline{\smallskip}
\end{tabular}
\end{center}
\label{tab:sources}
Notes. -- Col. (1) Cluster name; cols. (2), (3) and (4) cluster coordinates 
and redshift from the NASA/IPAC Extragalactic Database (NED); col. (5) 
angular to linear scale conversion.
$(a)$ spectroscopic redshift of the BCG from the Sloan Digital Sky 
Survey Data Release 9 (SDSS DR9). A redshift of $z=0.345$ was initially 
reported in the ROSAT Brightest Cluster Sample \citep{1998MNRAS.301..881E}, 
but later optical follow-up established that the cluster is at 
$z=0.3615$ \citep{1999MNRAS.306..857C}.
\end{table}


\section{Radio observations and data reduction}\label{sec:obs}

Here we only briefly give the relevant information on our cluster sample 
in which we searched for minihalos. More details on the sample selection 
will be presented in a future paper about correlations of various 
cluster properties. We selected all those 
clusters from the {\em Chandra} ACCEPT\footnote{Archive of Chandra Cluster Entropy Profile Tables.} 
sample \citep{2009ApJS..182...12C} with deep, pointed radio observations suitable to 
look for a central minihalo or place an interesting upper limit on it,
obtaining a sample of $\sim 100$ clusters. The radio information was 
obtained from the {\em VLA} archive and/or proprietary {\em GMRT} 
data, mostly from the {\em GMRT} Radio Halo Survey \citep[GRHS;][]{2007A&A...463..937V,2008A&A...484..327V} 
and its recent extension \citep[eGRHS;][]{2013A&A...557A..99K}.
We considered only clusters at $z\le 0.5$ and DEC$_{\rm J2000} > -30^{\circ}$, 
to ensure a good sampling of the $u-v$ plane of the radio observations,
necessary to properly image extended and complex radio sources such as 
minihalos. A minihalo has previously been reported for 
12 clusters of our sample. In this paper, we report detections
of new minihalos or candidates for 9 more clusters (listed in Table 1),
based on re-analysis of the {\em VLA} archival observations.
The {\em VLA} observations are summarized in Table 2.

The data were calibrated and reduced using the NRAO Astronomical Image Processing
System package (AIPS). Images were produced using the standard Fourier
transform deconvolution method. Self-calibration was applied to reduce
the effects of residual phase errors in the data and improve the quality
of the final images. Correction for the primary beam attenuation was
applied to the images using the task PBCOR in
AIPS. All flux densities are in the \cite{2013ApJS..204...19P}
scale and residual amplitude errors are within $5\%$ at all frequencies.



\begin{table*}[t]
\caption[]{Details of the {\em VLA} observations}
\begin{center}
\begin{tabular}{lrcccccrc}
\hline\noalign{\smallskip}
\hline\noalign{\smallskip}
Cluster name & Project & Array & Frequency & Bandwidth & Date &
Time  & FWHM, PA  &   rms    \\  
  &                  &            & (GHz)      &      (MHz)   &   &  (min)    &
  ($^{\prime \prime} \times^{\prime \prime}$, $^{\circ}$)\phantom{00}
  & ($\mu$Jy b$^{-1}$) \\
\noalign{\smallskip}
\hline\noalign{\smallskip}
MACS\,J0159.8$-$0849\dotfill & AE147 & B & 1.4 & 50 & 2002 Jul 15 & 48 & $5.7\times4.2$, $-17$ & 15 \\
                             & AE117 & A & 8.5 & 50 & 1998 Apr 12 & 19 & $0.3\times0.2$, $-28$ & 18 \\

MACS\,J0329.6$-$0211\dotfill  & AE142 & B & 1.4 & 50 & 2001 May 05&
48 & 4.7$\times$4.5, 0 & 40 \\

A\,478\dotfill  & AB1150 & A  & 1.4 & 25 & 2004 Nov 24 & 320 &
$1.3\times1.3$, $-12$ & 25 \\
           & AM938 & C & 1.4 & 50 & 2009 Jul 18 & 190 &
           $14.2\times11.8$, $-49$ & 40 \\
           &  AK685 & C&  4.9 & 50 & 2008 May 19 & 93 &
          $4.1\times3.8$, $-29$ & 15  \\

ZwCl\,3146\dotfill  & AB1190 & C & 4.9 & 50 & 2006 Dec 18 & 272 &
$4.3\times3.9$, 40 & 10 \\
             & AB1190 & C&  8.5 & 50 & 2006 Dec 18 & 102 &
             $2.5\times2.4$, $-34$ & 15 \\

 A\,1795\dotfill 
              & AO84 & A & 1.4 & 3 & 1989 Jan 18 & 176 &
              $1.3\times1.2$, 69 & 40 \\ 
              & AJ215 & C & 1.4 & 6 & 1992 Feb 25 & 130 & 14.1$\times$13.4, 40 & 150 \\
              
RX\,J1532.9+3021\dotfill  & AT0318 & A & 1.4 & 50 & 2006 Mar 15 &
72 &  $1.4\times1.1$, $-29$ & 12 \\
                                 & AT0318 & B &1.4 & 50 &
                                 2006 Sep 11 & 50 & $4.2\times3.7$,
                                 $-11$ & 15 \\
                                 & AT0318 & B&  0.3 & 6
                                     & 2006 Mar 15 & 118  & 
                                     $5.2\times4.8$, $-27$ & 500  \\
                                 & AE110 & C & 4.9 & 50 &
                                 1997 Jun 26 & 6 &  $4.8\times3.9$,
                                 $-75$ & 20 \\
                                  & AE117 & A & 8.5 & 50 & 1998 Apr 12
                                  & 11 & $0.2\times0.2$, 42 & 25 \\
                                  & AK633 & D & 8.5 &  50 & 2005 Apr 7
                                  & 5 &
                                  $8.1\times7.4$, $-73$ & 30 \\
                                  & AK633 & D & 22 & 50 & 2007 Apr 5&
                                  4 &  $2.9\times2.6$, $-75$ & 140  \\
 
A\,2204\dotfill  & AT0211 & A & 1.4 & 25 & 1998 Apr 23  & 123 &
$1.3\times1.3$, 1&  30 \\
             & S8398 & B & 1.4 & 50 & 2007 Nov 25 & 102 &
             $4.7\times4.2$, 38 & 30 \\

ZwCl\,1742.1+3306\dotfill  & AE130 & DnA & 1.4 & 6 & 1999 Jun 10 & 145 &
$1.5\times1.5$, 47 &  65 \\
                           
MACS\,J1931.8--2634\dotfill  & AT0318 & A & 1.4 & 50 & 2006 Apr 14 &
82 & $2.5\times1.2$, 3 &  30  \\
                                     & AT0318 & B & 1.4 &
                                     50 & 2006 Sep 09 & 49 &
                                     $8.4\times3.8$, 0 & 40  \\
                                     & AT0318 &A & 0.3 & 6.3
                                     & 2006 Apr 14 & 120 &
                                     $11.0\times4.5$, 1 &  1700  \\

\hline{\smallskip}
\end{tabular}
\end{center}
\label{tab:obs}
Notes. -- Col. (1) cluster name; cols. (2) and (3) {\em VLA} array configuration 
and project; cols. (4), (5) and (6) observing frequency, total bandwidth and date; 
col. (7) total time on source; col. (8) full-width half maximun (FWHM) and position
angle (PA) of the array; col. (9) image rms level ($1\sigma$).
\end{table*}



\begin{table}
\caption[]{List of the {\em Chandra} observations}
\label{tab:obs}
\begin{center}
\begin{tabular}{cccc}
\hline\noalign{\smallskip}
Cluster name & Detector & Observation ID & Exposure$^a$\\
                      &                   &    &(ksec ) \\
\noalign{\smallskip}
\hline\noalign{\smallskip}
MACS\,J0159.8--0849\dotfill & ACIS--I    & 3265, 6106, 9376  & 70\\
MACS\,J0329.6--0211\dotfill & ACIS--I & 3257, 3582, 6108  & 70  \\
A\,478\dotfill & ACIS--S & 1669 & 40 \\
ZwCl\,3146\dotfill & ACIS--I & 909, 9371 & 90 \\
A\,1795\dotfill & ACIS--S & 493, 494 & 40  \\
RX\,J1532.9+3021\dotfill &  ACIS--S, I, S & 1649, 1665, 14009 & \phantom{0}105$^b$  \\
A\,2204\dotfill & ACIS--S, I, I & 499, 6104, 7940 & 100  \\
ZwCl\,1742.1+3306\dotfill& ACIS--S &11708 & 45 \\ 
MACS\,J1931.8-2635\dotfill  & ACIS--I & 9382 & 100  \\
\hline{\smallskip}
\end{tabular}
\end{center}
\label{tab:xray}
Notes. -- $(a)$ total exposure time without filtering; $(b)$ total clean exposure from
\cite{2013arXiv1306.0907H}.
\end{table}



\begin{table*}[t]
\caption[]{Properties of the discrete radio sources}
\begin{center}
\begin{tabular}{cccccrccc}
\hline\noalign{\smallskip}
Cluster name  & source &  RA$_{\rm J2000}$ & DEC$_{\rm
  J2000}$ & $\nu$ &  S$_{\nu}$\phantom{000}  & $\alpha$ & Notes \\
             &             &     (h, m, s) &  ($^{\circ}$, $^{\prime}$,
             $^{\prime \prime}$)   &  (GHz)  &  (mJy)\phantom{00} & 
           & \\
\noalign{\smallskip}
\hline\noalign{\smallskip}
MACS\,J0159.8--0849\dotfill & S1 &  01 59 49.3 & $-08$ 49 59 & 1.4 & $35.0\pm1.8$ & $-0.55\pm0.04$  &  BCG \\
                            &    &             &             & 8.5 & $94.8\pm4.8$ &                 &   \\
                            & S2 &  01 59 50.9 & $-08$ 49 50 & 1.4 & $0.43\pm0.03$ &     $-$        & possible member galaxy$^a$ \\
&&&&&&&\\

MACS\,J0329.6--0211\dotfill & S1 & 03 29 41.6 & $-02$ 11 47 & 1.4 & $3.8\pm0.2$&  $-$  &  BCG  \\

&&&&&&&\\

A\,478\dotfill & S1 &  04 13 25.3 & +10 27 55 & 1.4 & $31.0\pm1.6$ & $0.92\pm0.06$ &  BCG \\
               &    &             &           & 4.9 &  $9.6\pm0.5$ &               &      \\

               & S2 & 04 13 31.8  & +10 28 39 & 1.4 &  $2.5\pm0.1$ &$1.03\pm0.06$  & member galaxy$^b$ \\
               &    &             &           & 4.9 & $0.68\pm0.04$&               &                                \\

               & S3 & 04 13 35.2  & +10 29 21 & 1.4 & $1.10\pm0.06$ &$-0.08\pm0.06$& no $z$ \\

               & S4 & 04 13 38.3  & +10 28 09 & 1.4 & $47.4\pm2.4$ & $0.93\pm0.05$ & member galaxy$^c$ \\
               &    &             &           & 4.9 & $14.6\pm0.7$ &               &                             \\
&&&&&&&\\   

ZwCl\,3146\dotfill&  S1 & 10 23 39.6 & +04 11 11 & 1.4 & $\sim 3.3 ^d$      & $0.67\pm0.10 ^e$ & BCG  \\
                  &     &            &           & 4.9 & $1.42\pm0.07$ &                  &              \\
                  &     &            &           & 8.5 & $0.98\pm0.03$ &                  &              \\

                  & S2  & 10 23 38.7 & +04 11 05 & 1.4 & $0.2 ^d$      & $-0.32\pm0.15 ^e$& background galaxy$^f$ \\
                  &     &            &           & 4.9 & $0.31\pm0.02$ &                  &         \\
                  &     &            &           & 8.5 & $0.37\pm0.02$ & && \\
&&&&&&&\\
A\,1795\dotfill   & S1  & 13 48 52.5 & +26 35 34 & 1.4 & $917\pm46^g$ &             & BCG    \\


                  & S2  & 13 48 45.5 & +26 35 24 & 1.4 & $5.1\pm0.3$  &               & background galaxy$^h$ \\

&&&&&&&\\
RX\,J1532.9+3021\dotfill & S1  & 15 32 53.8 &  +30 20 59 & 0.3  &
$26.3\pm1.4$ & $0.43\pm0.02$ & BCG \\
                                 &     &  &  & \phantom{0}0.6$^i$   & $20.3\pm1.0$ & &
                                  \\
                                 &  & &     & 1.4  & $15.7\pm0.8$ &  &
                                  \\
                                &     & &  & 4.9 & $8.8\pm0.4$ & &\\
                               &&&       & 8.5 & $7.0\pm0.4$ && 
                                \\ 
                                & & && 22 & $4.2\pm0.2$ & & \\

&&&&&&&\\
A\,2204\dotfill &  S1  & 16 32 47.0 & +05 34 35 & 1.4 & $58.9\pm2.9$  & $-$   & BCG \\

                &  S2  &  16 32 47.0 & +05 34 41 & 1.4 & $1.6\pm0.1$ & $-$ & no $z$ \\
&&&&&&&\\
ZwCl\,1742.1+3306\dotfill & S1 & 17 44 14.5 & 32 59 29 & 1.4 & $68.8\pm3.4$ & $-$  &  BCG \\
                          & S2 & 17 44 17.1 & 32 59 14 & 1.4 & $0.94\pm0.05$ & $-$ & no $z$\\
&&&&&&&\\
MACS\,J1931.8--2634\dotfill & S1 & &  &1.4 & $11.6\pm0.05$ & & BCG \\
                                   & S2 &  & &1.4 & $2.5\pm0.1$ & &  no
                                   opt. id.\\
\hline{\smallskip}
\end{tabular}
\end{center}
\label{tab:sources}
Notes. -- Col. (1) cluster name; 
col. (2) radio source; 
cols. (3) and (4) radio coordinates; 
col. (5) frequency; 
col. (6) flux density; 
col. (7) spectral index; 
col. (8) notes on the optical identification. 
$(a)$ $z_{\rm phot}=0.44\pm0.04$ from SDSS DR9; 
$(b)$ $z=0.0894\pm0.0003$ \citep{1990ApJS...74....1Z};
$(c)$ $z=0.0934\pm0.0002$ \citep{1990ApJS...74....1Z};
$(d)$ expected value based on $\alpha=0.67$ between 4.9 GHz and 8.5 GHz; 
$(e)$ spectral index between 4.9 GHz and 8.5 GHz; 
$(f)$ $z_{\rm phot}=0.34\pm0.04$ from SDSS DR9; 
$(g)$ measured on the FIRST image; a flux density of $890\pm45$ mJy 
is measured on the A-array image (Fig.~\ref{fig:a1795}(b)) and 
$960\pm48$ mJy on the C-array image (Fig.~\ref{fig:a1795}(a)); 
$(h)$ $z_{\rm phot}=0.57\pm0.08$ from SDSS DR9; 
$(i)$: data from \cite{2013A&A...557A..99K}.
\end{table*}




\begin{table*}[t]
\caption[]{Properties of minihalos and their cluster hosts} 
\begin{center}
\begin{tabular}{lcccccccc}
\hline\noalign{\smallskip}
Cluster  name &  S$_{\rm MH, \,1.4 \, GHz}$ & \phantom{0}$P_{\rm MH, \,1.4 \, GHz}$ & $R_{\rm MH}$ & \phantom{0}$P_{\rm BCG, \,1.4 \, GHz}$ &  $kT_{\rm ncc}$ & $M_{500}$ &  Ref.  & notes\\
              & (mJy)  & ($10^{24}$ W Hz$^{-1}$) & (kpc) & ($10^{24}$ W Hz$^{-1}$)  & (keV) & $10^{14}$ $M_{\odot}$ & &   \\
\noalign{\smallskip}
\hline\noalign{\smallskip}
New detections &&&&& \\
&&&&&\\
MACS\,J0159.8--0849\dotfill & $2.4\pm0.2$ & $1.40\pm0.14$ & 90 & $20.3\pm0.1$  &$9.36^{+0.77}_{-0.67}$ & $6.88^{+0.90}_{-0.98}$ & 1,5 & candidate \\
MACS\,J0329.6--0211\dotfill & $3.8\pm0.4$ & $2.84\pm0.30$ & \phantom{0}70$^a$ & $2.8\pm0.1$ & $6.44^{+0.50}_{-0.45}$ & $-$\phantom{000} & 1,5 & candidate\\

A\,478\dotfill & $16.6\pm3.0$ & $0.32\pm0.06$ & 160$^b$ & $0.60\pm0.03$ &$7.27^{+0.26}_{-0.25}$  &  $7.06^{+0.35}_{-0.36}$ & 1,5 &\\
ZwCl\,3146\dotfill & $\sim 5.2^c$ & $\sim 1.39$ & \phantom{0}90$^d$ & $\sim 0.8^e$ & $7.46^{+0.32}_{-0.30}$ & $-$\phantom{000} & 1,5 & \\

A\,1795\dotfill & $85.0\pm4.9$ & $0.79\pm0.05$ & \phantom{0}100$^f$ &$8.5\pm0.4$ &$6.05^{+0.15}_{-0.15}$ & $4.54^{+0.21}_{-0.21}$ & 1,5& candidate \\
RX\,J1532.9+3021\dotfill & $7.5\pm0.4$ & $3.35\pm0.17$ & 100  & $7.0\pm0.4$ &$6.06^{+0.43}_{-0.39}$ & $-$\phantom{000} & 1,5 & \\

A\,2204\dotfill & $8.6\pm0.9$ & $0.54\pm0.05$ & 50 & $3.7\pm0.2$ & $9.35^{+0.43}_{-0.41}$  & $7.96^{+0.37}_{-0.38}$&1,5 &  \\
ZwCl\,1742.1+3306\dotfill & $13.8\pm0.8$ & $0.20\pm0.01$ & 40 & $0.97\pm0.05$ & $5.23^{+0.84}_{-0.73}$  & $6.44^{+0.71}_{-0.76}$ &1,6  & uncertain$^f$\\
MACS\,J1931.8--2634\dotfill & $47.9\pm2.8$ & $20.0\pm1.2$ & 100 & $4.9\pm0.2$ & $6.85^{+0.73}_{-0.61}$ & $6.19^{+0.77}_{-0.83}$ & 1,5& uncertain$^g$ \\ 
\hline{\smallskip}
&&&&&\\
Previously known minihalos  &&&&& \\
&&&&& \\
Perseus\dotfill    &  $3020\pm153$\phantom{0}  & $2.18\pm0.11$ & 130 & $13.4\pm0.1^h$  & $6.42^{+0.06}_{-0.06}$ *       & $-$\phantom{000} & 2,7 \\
A\,1835\dotfill    &  $6.1\pm1.3$\phantom{0}     & $1.19\pm0.25$ &  \phantom{0}240$^i$  & $6.3\pm0.3$ & $9.77^{+0.55}_{-0.51}$  &$8.46^{+0.55}_{-0.57}$ & 3,5 \\
Ophiuchus\dotfill  & $83.4\pm6.6^j$ & $0.15\pm0.02$ & 250 & $0.064\pm0.003$&$10.25^{+0.18}_{-0.22}$ *  & $-$\phantom{000}  & 3,7 \\
A\,2029\dotfill    &$19.5\pm2.5$\phantom{0}    & $0.28\pm0.04$ & \phantom{0}270$^i$ & $7.4\pm0.4$ & $8.22^{+0.31}_{-0.30}$  & $6.82^{+0.24}_{-0.25}$ & 3,5 \\
A\,2390\dotfill    & $28.3\pm4.3$\phantom{0}    & $4.46\pm0.67$ & \phantom{0}250$^i$ & $33.2\pm1.7$  &$10.85^{+0.34}_{-0.31}$\phantom{0} &$9.48^{+0.41}_{-0.42}$ & 3,5 \\
RBS\,797\dotfill   & $5.2\pm0.6^k$  & $2.20\pm0.24$ & 120 & $7.6\pm0.4$ &$7.63^{+0.94}_{-0.77}$ &$6.27^{+0.63}_{-0.66}$  & 3,5  & \\ 
RXC\,J1504.1--0248\dotfill&  $20.0\pm1.0$\phantom{0} & $2.70\pm0.14$ & 140 & $5.7\pm0.3$ & $8.02^{+0.26}_{-0.25}$ &$6.98^{+0.57}_{-0.60}$ & 4,5 \\
RX\,J1347.5--1145\dotfill & $34.1\pm2.3$\phantom{0} & $25.59\pm1.79$ & 320 & $22.7\pm1.1$ & $15.12^{+1.03}_{-0.86}$\phantom{0} &$10.61^{+0.74}_{-0.77}$ & 3,5\\
RX\,J1720.1+2638\dotfill & $72.0\pm4.4$\phantom{0} & $5.33\pm0.32$ & 140 & $0.50\pm0.02$ & $6.33^{+0.29}_{-0.25}$ & $6.34^{+0.38}_{-0.40}$ & 3,5 \\
MS\,1455.0+2232\dotfill &  $8.5\pm1.1$\phantom{0} & $1.75\pm0.23$ & 120 & $0.96\pm0.05$ & $4.82^{+0.14}_{-0.13}$ &$-$\phantom{000} & 3,5\\
2A\,0335+096\dotfill &   $21.1\pm2.1^k$ & $0.059\pm0.006$ & 70 & $0.058\pm0.003$& $3.53^{+0.10}_{-0.13}$  & $2.27^{+0.24}_{-0.25}$ & 3,8\\
A\,2626\dotfill & $18.0\pm1.8$ & $0.14\pm0.01$ & 30 & $0.129\pm0.006$ & $3.29$ *  & $-$ &  9,10 & uncertain$^l$ \\
\hline{\smallskip}
\end{tabular}
\end{center}
\label{tab:halos}
Notes. -- Col. (1) Cluster name; 
cols. (2) and (3) minihalo flux density 
(after subtraction of the flux density of the embedded discrete radio sources) 
and corresponding radio power at 1.4 GHz; 
col. (4) average radius of the minihalo, defined as in Eq.~\ref{eq:r}; 
col. (5) radio power at 1.4 GHz of the BCG;
col. (6) cluster temperature with excision of the central cool core, 
except for those values marked with *; 
col. (7) total cluster mass within $R_{500}$ from \cite{2013arXiv1303.5089P};
col (8) references to radio flux density and cluster temperature: 
(1) this work, 
(2) Sijbring (1993), 
(3) Giacintucci et al. (in preparation), 
(4) \cite{2011A&A...525L..10G},
(5) \cite{2008ApJ...682..821C},
(6) \cite{2007A&A...466..805C},
(7) \cite{2002A&A...383..773I},
(8) \cite{2010A&A...513A..37H},
(9) \cite{2013MNRAS.436L..84G}
(10) \cite{2009ApJS..182...12C};
$(a)$ the minihalo may be more extended (see gray-scale image in 
Fig.~\ref{fig:macsj0329}(a));
$(b)$ uncertain due to the presence of the HT source S4 (Fig.~\ref{fig:a478}(c));
$(c)$ estimated from the NVSS image, after subtraction of 
the expected flux densities of the embedded radio galaxies (see \S~\ref{sec:3146});
$(d)$ measured on the 4.9 GHz image;
$(e)$ expected value, based on the source spectral index between 4.9 GHz and 8.5 GHz;
$(f)$ the minihalo may be more extended (see gray scale image in Fig.~\ref{fig:a1795}{\em a});
$(g)$ the classification as a minihalo is uncertain (see \S~\ref{sec:1931});
$(h)$ from NVSS;
$(i)$ uncertain due to the presence of many discrete radio galaxies in the area covered by the minihalo emission;
$(j)$ a flux density of 106.4$^{+10.4}_{-8.9}$ mJy has been measured by 
\cite{2009A&A...499..679M} using an exponential model to fit the surface 
brightness profile of the minihalo; a flux density of $85\pm3$ mJy 
is reported by Murgia et al. within a radius of $\sim 230$ kpc, which is
consistent with our measurement within a radius of $\sim 280$ kpc;
$(k)$ the contribution of the emission associated with the X-ray cavities 
has been subtracted out;
$(l)$ the extended emission seen in the core \citep[][and references therein]{2013arXiv1308.5825G} 
may be associated with the central radio galaxy rather than being a minihalo.
\end{table*}


\section{Radio and X-ray data}\label{sec:images}

The {\em VLA} setup for our clusters is presented in Table 2.
For almost all clusters, it is possible to image the central radio emission 
at different angular resolutions. The higher-resolution images are used to 
identify the radio source associated with the brightest cluster galaxy (BCG) 
and other possible radio galaxies in (or projected onto) the cluster core 
region. Low-resolution images are then used to map the diffuse radio emission. 
The radio images are compared to the X-ray {\em Chandra} images to search for 
possible indications of a spatial correlation of the diffuse radio emission 
with the ICM substructures (e.g, cold fronts, surface brighntess edges, 
spiral-like features). A detailed study of any radio/X-ray surface brightness 
correlation found here will be presented in future papers. Table 3 lists 
the {\em Chandra} observations used to produce the X-ray images. Since we 
are interested only in a qualitative analysis of the core structure, 
for most of our clusters we extracted images directly from the standard event files 
produced and archived by the {\em Chandra} X-ray Center, without additional 
data filtering, background subtraction or exporure map correction. 
For RX\,J1532.9+3021, we use an exposure-map corrected image that combines 
three available observations for a total clean exposure of 105 ks 
\citep[courtesy of J. Hlavacek-Larrondo; for details see][]{2013ApJ...777..163H}.

We summarize the position and flux densities of the radio galaxies in
each cluster image, their total spectral index and optical identification in 
Table 4. In Table 5, we summarize the radio properties of the new minihalos 
and candidates presented in this paper. For all clusters, we determined the 
position and flux density of the unresolved radio galaxies by fitting the 
sources with a Gaussian model (task JMFIT in AIPS). 

For extended radio galaxies, we measured the total flux 
density by integration within the $+3\sigma$ surface brightness contour using 
the task TVSTAT on the final images. For the minihalos, we measured the
total flux in circular regions of a radius that progressively increased
from the radius encompassing the $+3\sigma$ isocontour until the integrated
flux density reached saturation. For most minihalos, more than 95\% of the
flux is contained already within the $+3\sigma$-isocontour radius.

Following \cite{2013arXiv1306.4379C}, we estimated the error on the minihalo flux density $S_{\rm MH}$ as  

\begin{equation}
\sigma_{S_{\rm MH}} = \sqrt{(\sigma_{\rm cal} S_{\rm MH})^2 + (rms \sqrt{N_{\rm beam}})^2 + \sigma_{\rm sub}^2}
\end{equation}

\noindent which takes into account the uncertainty on the flux 
density scale ($\sigma_{\rm cal}\sim 5\%$), the image rms level weighted by the 
number of beams in the minihalo region ($N_{\rm beam}$), and the uncertainty 
$\sigma_{\rm sub}$ in the subtraction of the flux density of the 
embedded discrete radio sources from the total flux density measured in 
the image. This latter is estimated as 

\begin{equation}
\sigma_{sub}^2=\sum_{i=1}^{N} (I_{MH,i} \times A_{s,i})^2
\end{equation}

\noindent where $I_{H,i}$ is the average surface brightness 
of the minihalo in the region of the $i-$th radio source, 
occupying an area $A_{s,i}$.

Because of the irregular and non-spherical morphology of some minihalos, 
we estimated the average radius of the diffuse emission as

\begin{equation} 
R_{\rm MH} = \sqrt{R_{\rm max} \times R_{\rm min}}
\end{equation}\label{eq:r}

\noindent where $R_{\rm max}$ and $R_{\rm min}$ are the maximum and minimum radius 
as derived from the $+3\sigma$ isocontour in the images \citep{2007MNRAS.378.1565C}.
However, the sizes determined in this way may be affected by the different 
signal-to-noise ratios of the images and should be therefore considered as a lower limit for the real sizes.
A more appropriate definition of the size, indipendent on the sensitivity of the observations, should 
be used for quantitative studies of the minihalo properties (e.g., Murgia et al. 2009). 
We will adress this in a future paper by analysing the surface brightness radial profiles of all minihalos, 
including the new detections presented here.

In Table 5, we also provide the list of the clusters hosting 
known minihalos. For consistency with the present work,  
we have re-analyzed the existing {\em VLA} archival observations of
all these systems (but Perseus and A\,2626, for which we
use the literature information as reported in the table) 
and re-derived the minihalo flux densities and sizes 
following the procedure described above (Giacintucci et al. in prep.). 
We note that our list does not include the clusters A\,2142 and 
MRC\,0116+111 in which minihalos have been reported in the literature.
Recent {\em Green Bank Telescope} ({\em GBT}) observations 
at 1.4 GHz \citep[Farnsworth et al. submitted;][]{2013A&A...556A..44R}
and {\em GMRT} observations at lower frequencies 
(Venturi et al. in preparation) reveal that the central diffuse source is 
A\,2142 is far more extended than the $\sim 200$ kpc minihalo imaged by 
\cite{2000NewA....5..335G}, covering a scale of the order of the Mpc, and, 
thus, falling into the category of ``giant'' radio halos \citep[e.g.,][]{2012A&ARv..20...54F}. 
No X-ray information is avaiable for MRC\,0116+111, thus the classification of the source as 
a cluster minihalo \citep{2009MNRAS.399..601B} is very uncertain \citep{2012A&ARv..20...54F}.

Besides the radio properties of the minihalos, Table 5 provides 
values of radio power of the BCG, cluster X-ray temperature and total mass 
that will be used in \S\ref{sec:disc} for the discussion of our results. 
$M_{500}$, the cluster mass within $R_{500}$\footnote{$R_{500}$ is the radius 
corresponding to a total density contrast $500\rho_c(z)$, $\rho_c(z)$ being 
the critical density of the Universe at the cluster redshift.}, is estimated 
from the Sunyaev--Zeldovich (SZ) effect measured by {\em Planck} as described 
in Planck collaboration (2013). 

%
%
\begin{figure*}
\centering
\includegraphics[scale=1.3]{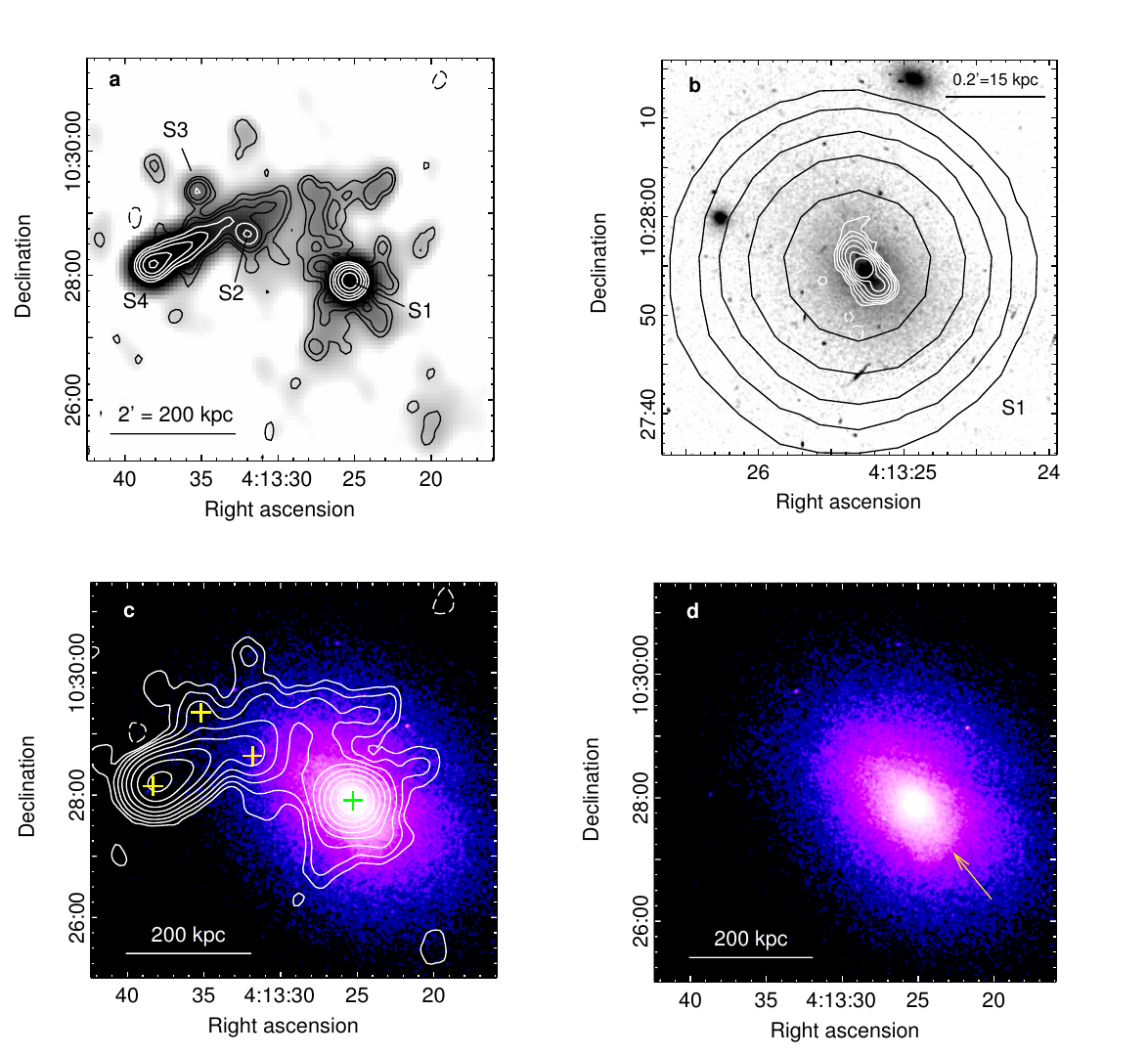}
\smallskip
\caption{A\,478. In all images, radio contours are $-1$ (dashed),
  $1, 2, 4, 8, 16, ... \times3\sigma$. (a) {\em VLA} 1.4 GHz C--array image 
(white and black contours), restored with a $15^{\prime\prime}$ circular beam. 
The r.m.s. noise is $1\sigma=40 \, \mu$Jy beam$^{-1}$. The image is also 
shown in gray scale, smoothed with a Gaussian with a radius of 5 pixels 
($20^{\prime\prime}$). The individuals radio galaxies, listed in Table 4,
are labelled. (b) {\em VLA} 1.4 GHz A--array contours of S1 (white), restored with a 
$1^{\prime\prime}$ circular beam, overlaid on the optical {\em HST} WFPC2 image 
(gray scale). The r.m.s. noise is $1\sigma=25\, \mu$Jy beam$^{-1}$. 
The C-array contours of S1 are reported in black (same as white contours in 
panel (a)). (c) Low-resolution {\em VLA} C--array contours at 1.4 GHz of the
 minihalo and radio galaxies (green and yellow crosses), overlaid on the 
{\em Chandra} image in the 0.5--4 keV band. 
The radio image has been restored with a $30^{\prime\prime}$ circular
beam. The r.m.s. noise is $1\sigma=50 \, \mu$Jy beam$^{-1}$.
 (d): {\em Chandra} image, same as in (c). The arrow indicates the 
position of a cold front.} 
\label{fig:a478}
\end{figure*}
%
%

\section{New radio minihalos}

Our reanalysis of the {\em VLA} data of the clusters 
A\,478, RX\,J1532.9+3021, ZwCl\,3146 and A\,2204 led to the detection of large-scale 
diffuse emission in their cool cores, which we classify as minihalos.
In the following sections, we describe these newly discovered minihalos 
and summarize their properties.

\subsection{A\,478}\label{sec:mh}

A\,478 is a nearby ($z=0.09$) cool-core cluster with a remarkably 
relaxed X-ray morphology and symmetric temperature distribution on 
large scales \citep[e.g.,][]{2008A&A...479..307B}. In the cool core, a 
high-resolution {\em Chandra} image shows significant substructure, 
resulting from the interaction of the central active radio galaxy with 
the surrounding ICM \citep{2003ApJ...587..619S}. Two small X-ray
cavities, located within the central $r\sim 9^{\prime\prime}$
($\sim15$ kpc), are partially filled by the radio galaxy lobes at 1.4
GHz. With a size of only $\sim$4 kpc, these are among the smallest 
cavities found in cluster cores.

{\em VLA} 1.4 GHz images of A\,478 at several resolutions 
are presented in Fig.~1. The C-configuration image in panel (a) 
reveals the presence of a diffuse, low surface brightness radio source,
which we classify as a minihalo. The minihalo encompasses 
the central bright radio galaxy (S1), unresolved at this resolution 
(FWHM=$15^{\prime\prime}$). At higher resolution, S1 has a double-lobe 
morphology with a size of $\sim$13 kpc, as seen in Fig.~1(b), which 
shows the A-array image (white contours) overlaid on the {\em HST} 
Wide-Field Planetary Camera 2 (WFPC2) image (gray scale). We also 
report in black the C-array contours of S1 from Fig.~1(a) to 
highlight the small size of the radio galaxy compared to the much 
larger scale of the surrounding minihalo. 

East of the minihalo, a head-tail source (S4), which is a cluster member (Table 4), 
extends for $\sim 200$ kpc, encompassing the unresolved source S2. 
This latter is another cluster member (Table 4) and possesses a tailed 
morphology at arcsecond resolution (image not shown here), with both jets bent toward 
north-east, i.e., almost perpendicular to the tail of S4. The unresolved 
source S3, north of S2 and S4, has no optical identification and is most 
likely a background source.

We produced a low angular resolution image to highlight the extended 
emission associated with the minihalo. The image is shown in Fig.~1(c), 
overlaid on the {\em Chandra} image. The minihalo appears more extended 
toward north-east ($R_{\rm max}\sim 180$ kpc), following a similar elongation 
of the X-ray surface brightness; its extent is $R_{\rm min}\sim 150$ kpc 
in the other directions. However, to the NE, the minihalo emission 
partially blends with the tail of S4 and it is difficult to 
determine the boundaries of the two structures. 

We measured the flux densities of S1, S2 and S3 on the A 
configuration image; consistent values were measured 
at lower resolutions. The emission 
associated with the extended source S4 was measured 
on the C configuration image (Fig.~\ref{fig:a478}(a)). 
S1, S2 and S4 are also
detected at 4.9 GHz (Table 2; image not shown here).
All flux densities and spectral indeces are summarized in 
Table 4. 

We subtracted the contribution of the radio galaxies S1-S4
from the total emission in Fig.~\ref{fig:a478}(c)
and measured a flux density of $16.6\pm3.0$ mJy 
for the minihalo, where the large error ($\sim 18\%$; Eq.~1) 
reflects the difficulty of subtracting the head-tail S4. 
The corresponding radio power is $P_{\rm  1.4 \, GHz} 
= (3.2\pm0.6) \times10^{23}$ W Hz$^{-1}$.

A cold front has been detected in the {\em Chandra} 
image at $\sim 60$ kpc south-west of the center (Markevitch et al. 2003).
Its position, marked by the arrow in Fig.~\ref{fig:a478}(d),
is roughly coincident with the southwestern boundary of the minihalo,
suggesting that the diffuse radio emission may be confined 
here by the front, as seen in other minihalos 
\citep[][Giacintucci et al. in prep.]{2008ApJ...675L...9M,2013ApJ...762...78Z}.

%
%
\begin{figure*}
\centering
\includegraphics[scale=1.25]{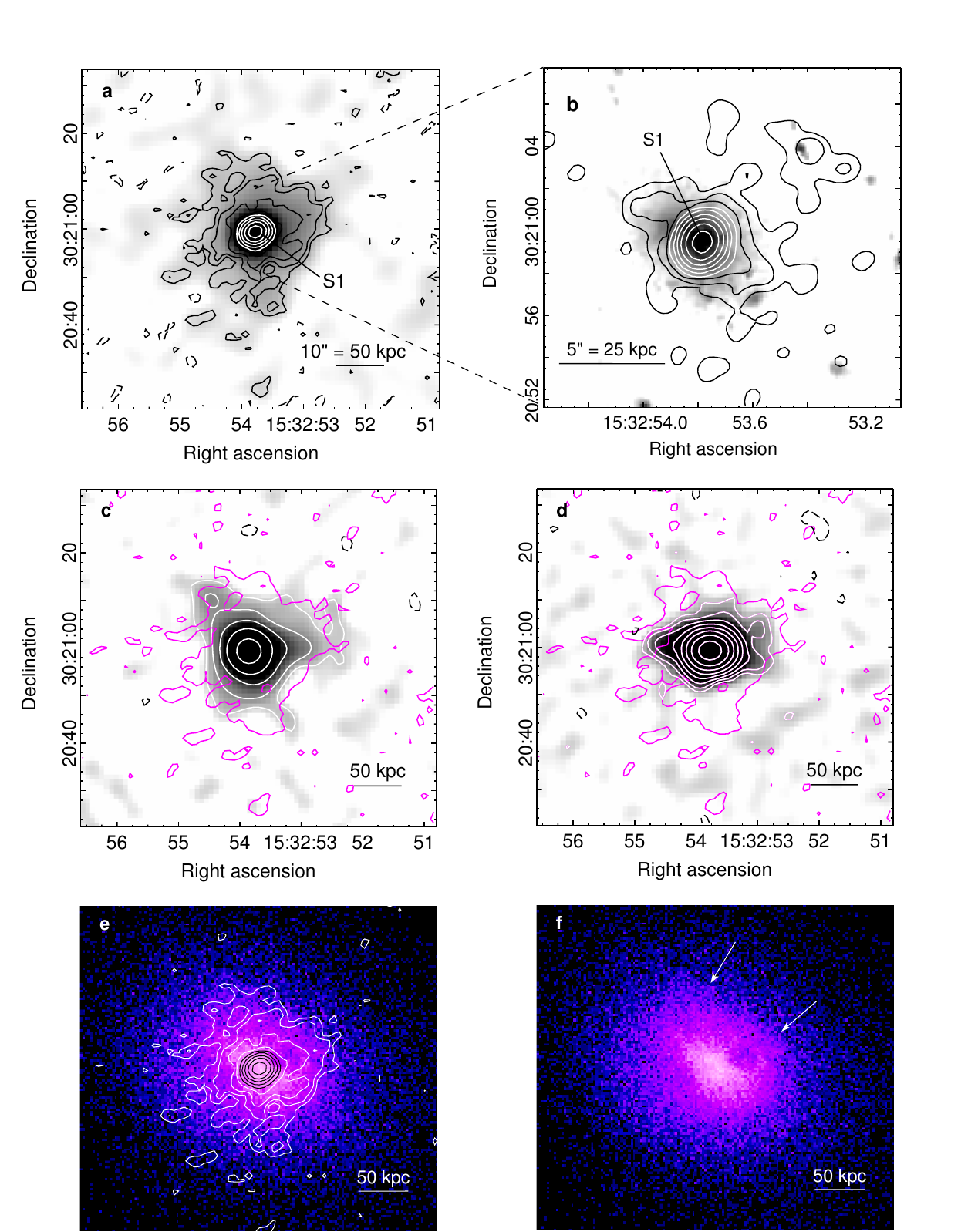}
\smallskip
\caption{RX\,J1532.9+3021. In all images, radio contours are $-1$ (dashed), 
$1, 2, 4, 8, 16, ... \times 3\sigma$. (a) {\em VLA} 1.4 GHz contours (black and white) 
from the combined A+B array. The beam is $3.4^{\prime\prime}\times2.9^{\prime\prime}$, in p.a. $-65^{\circ}$
and $1\sigma=10 \, \mu$Jy beam$^{-1}$. The image is also shown in gray scale, smoothed with a Gaussian 
with a radius of 4 pixels ($3^{\prime\prime}$). S1 is the central radio galaxy (see panel (b) and Table 4).
 (b) {\em VLA} A--array 1.4 GHz contours (black and white), overlaid on 
the {\em HST} WFPC2 image (gray scale).
The beam is $1.4^{\prime\prime}\times1.1^{\prime\prime}$, in
p.a. $-29^{\circ}$ and $1\sigma=12$ $\mu$Jy beam$^{-1}$. (c)  
{\em VLA} A-array image at 325 MHz (gray scale and white contours). 
The beam is $7.0^{\prime\prime}\times6.5^{\prime\prime}$, in
p.a. $-37^{\circ}$ and $1\sigma=400$ $\mu$Jy beam$^{-1}$. The
lowest contour at 1.4 GHz from panel (a) is shown in magenta. (d)
{\em VLA} C-array image at 4.9 GHz (gray scale and white contours).
 The beam is $6.0^{\prime\prime}\times4.9^{\prime\prime}$, in
 p.a., $90^{\circ}$ and $1\sigma=15$ $\mu$Jy beam$^{-1}$. The
lowest contour at 1.4 GHz from panel (a) is shown in magenta.
(e) {\em VLA} 1.4 GHz contours (same as (a)), overlaid on the 
{\em Chandra} image in the 0.5-7 keV band (from Hlavacek-Larrondo et al. 2013).
The image has the same physical size as panel (a).
(f) {\em Chandra} image, same as in (e). Arrows indicate the position 
of a cold front. A prominent cavity is also visible west of the cluster center
(see Hlavacek-Larrondo et al. 2013).}
\label{fig:rxcj1532}
\end{figure*}
%
%

\vspace{1cm}\subsection{RX\,J1532.9+3021}\label{sec:1532}

RX\,J1532.9+3021 is an X-ray luminous 
cluster at $z=0.362$. It has a mean temperature of $\sim 6$ keV (Table 5) 
and one of the most luminous cool cores
known \citep{2013ApJ...777..163H}.
Recent {\em Chandra} data confirmed the presence
of an X-ray cavity west of the cluster 
center and revealed a second, less prominent, cavity on 
the opposite side. A cold front was also reported at 
a radius of 65 kpc, partially coincident with the edge of
the western cavity \citep[][see also Fig.~\ref{fig:rxcj1532}(f)]{2013ApJ...777..163H}.

The {\em VLA} observations analyzed in this paper reveal a prominent
minihalo in the cluster core.
Fig.~\ref{fig:rxcj1532}(a) shows the 1.4 GHz image of 
the minihalo from the combined A+B array data set 
($\sim 3^{\prime\prime}$ resolution). The diffuse source is relatively
round in shape, with a radius of
$\sim$100 kpc, and totally encloses the central unresolved 
source S1, associated with the BCG. Fig.~\ref{fig:rxcj1532}(b) shows
an optical {\em HST} image of the BCG with an 
overlay of the 1.4 GHz A-configuration contours of the source S1 
(unresolved) and some of the surrounding diffuse 
emission associated with the minihalo.

We also obtained images from the archival {\em VLA} observations 
at 325 MHz in A configuration and at 4.9 GHz in C configuration
(shown in Fig.~\ref{fig:rxcj1532}(c,d)), 
as well as at 8.5 GHz (A and D configurations; not shown) 
and 22 GHz (D configuration; not shown).
Details of all these observations are given in Table 2.
The 325 MHz image clearly shows diffuse emission around S1 on a spatial scale that is 
similar to the extent of the minihalo at 1.4 GHz
(overlaid in magenta). At 4.9 GHz only the innermost and brightest
portion of the minihalo is detected. The minihalo
is not detected in the images at higher frequencies
where only the source S1 is detected. This source is still unresolved 
at the resolution of $0.2^{\prime\prime}$ of the 8.5 
GHz A-array image, which implies that its linear size $<1$ kpc.

The detection of a minihalo in this cluster was also reported by \cite{2013ApJ...777..163H}, who presented similar {\em VLA} images 
at 1.4 GHz and 325 MHz from the same data sets and a 
spectral index index image between these two frequencies. 
The cluster was also observed with the {\em GMRT} at 610 MHz 
as part of the GRHS. A re-analysis
of the 610 MHz observations, aimed to image the newly discovered minihalo,
is presented by \cite{2013A&A...557A..99K}. The morphology of the diffuse
source at 610 MHz is similar in shape and size to the minihalo structure seen 
at 1.4 GHz (see their Fig.~2).

The flux densities measured for S1 at all frequencies, including the
610 MHz value from Kale et al (2013), are reported in Table \ref{tab:sources}. 
The source has a spectral index $\alpha=0.43\pm0.02$ between 325 MHz and 22 GHz.
After the subtraction of S1, the minihalo has a flux density 
of $7.5\pm0.4$ mJy at 1.4 GHz, which corresponds to a radio power 
of $(3.4\pm0.2)\times10^{24}$ W Hz$^{-1}$ (Table 5).
Its flux density is $33.5\pm4.4$ mJy at 325 MHz, $16.0\pm0.8$ mJy
at 610 MHz and $1.3\pm0.2$ mJy at 4.9 GHz (all excluding S1).
The integrated spectrum of the minihalo, 
based on these measurements, is shown in Fig.~\ref{fig:rxcj1532_sp} and discussed in 
\S \ref{sec:alpha}. The total spectral index 
is $\alpha=1.20\pm0.07$.

The minihalo image is overlaid on the {\em Chandra} image in 
Fig.~\ref{fig:rxcj1532}(e). The position of the 
cold front is indicated by the arrows in panel (f).
As noticed by \cite{2013ApJ...777..163H}, 
the minihalo emission is 
apparently contained within the region delineated by the front. 
Part of the extended emission is spatially coincident with a
prominent X-ray cavity to the west of S1 (also visible in panel (f))
and with a second, weaker cavity, which has been recently found by 
Hlavacek--Larrondo et al. (2013). This indicates that a pair of radio 
lobes may be superposed to the larger-scale minihalo emission
(see also Hlavacek--Larrondo et al. 2013).

In the field of RX\,J1532.9+3021, we detected a complex 
radio source, possibly associated with a nearby galaxy cluster.
The radio images and description of this source are given in Appendix A.

%
%

\begin{figure*}[t]
\centering
\includegraphics[scale=1.2]{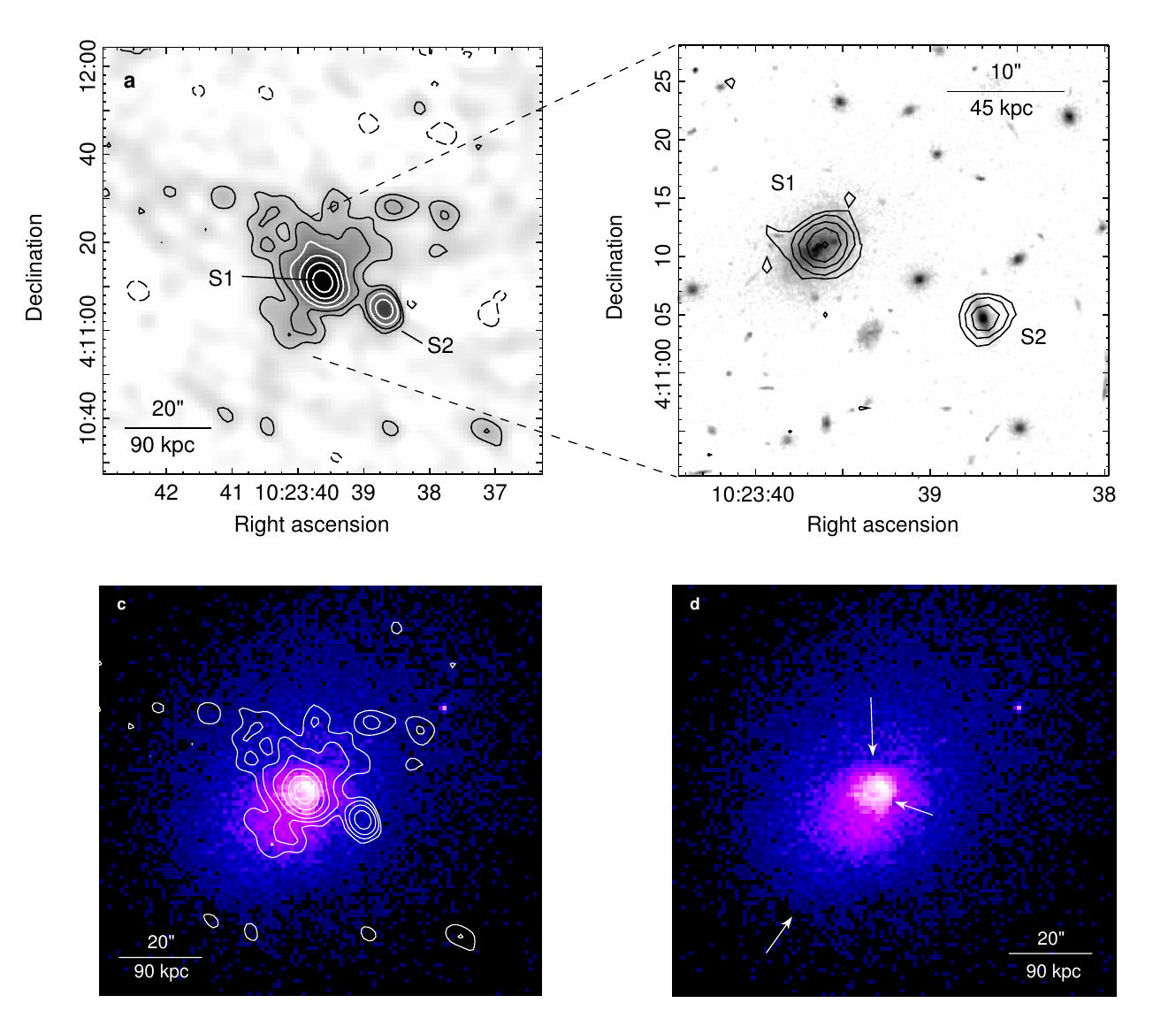}
\smallskip
\caption{ZwCl\,3146. (a) {\em VLA} C--array contours at 4.9
  GHz (black and white) at the resolution of
$5.7^{\prime\prime}\times4.6^{\prime\prime}$, in p.a. $27^{\circ}$. 
The r.m.s. noise is $1\sigma=10 \, \mu$Jy beam$^{-1}$.
Contours are $-1$ (dashed), $1,2, 4, 8, 16, ... \times 3\sigma$. 
The radio galaxies are labelled (see panel(b) and Table 4). 
The image is also shown in gray scale, smoothed with a Gaussian with a 
radius of 3 pixels ($3^{\prime\prime}$).
(b) {\em VLA} C--array contours at 8.5 GHz, overlaid on 
the {\em HST} WFPC2 image (gray scale). The restoring beam is 
$2.5^{\prime\prime}\times2.4^{\prime\prime}$, in p.a. $-34^{\circ}$. 
The r.m.s. noise is $1\sigma=15$ $\mu$Jy beam$^{-1}$. 
Contours are $1,2, 4, 8, 16, ... \times 3\sigma$. No levels
at $-3\sigma$ are present in the portion of the image shown.
(c) 4.9 GHz radio contours (same as in (a)) overlaid on the 
{\em Chandra} image in the 0.4-5 keV band. The image has the same
physical scale as panel (a). {\em d}: {\em Chandra} image, same 
as in (c). Arrows show the position of the cold fronts.}
\label{fig:zw3146}
\end{figure*}
%
%

%
%
\begin{figure*}
\centering
\includegraphics[scale=1.3]{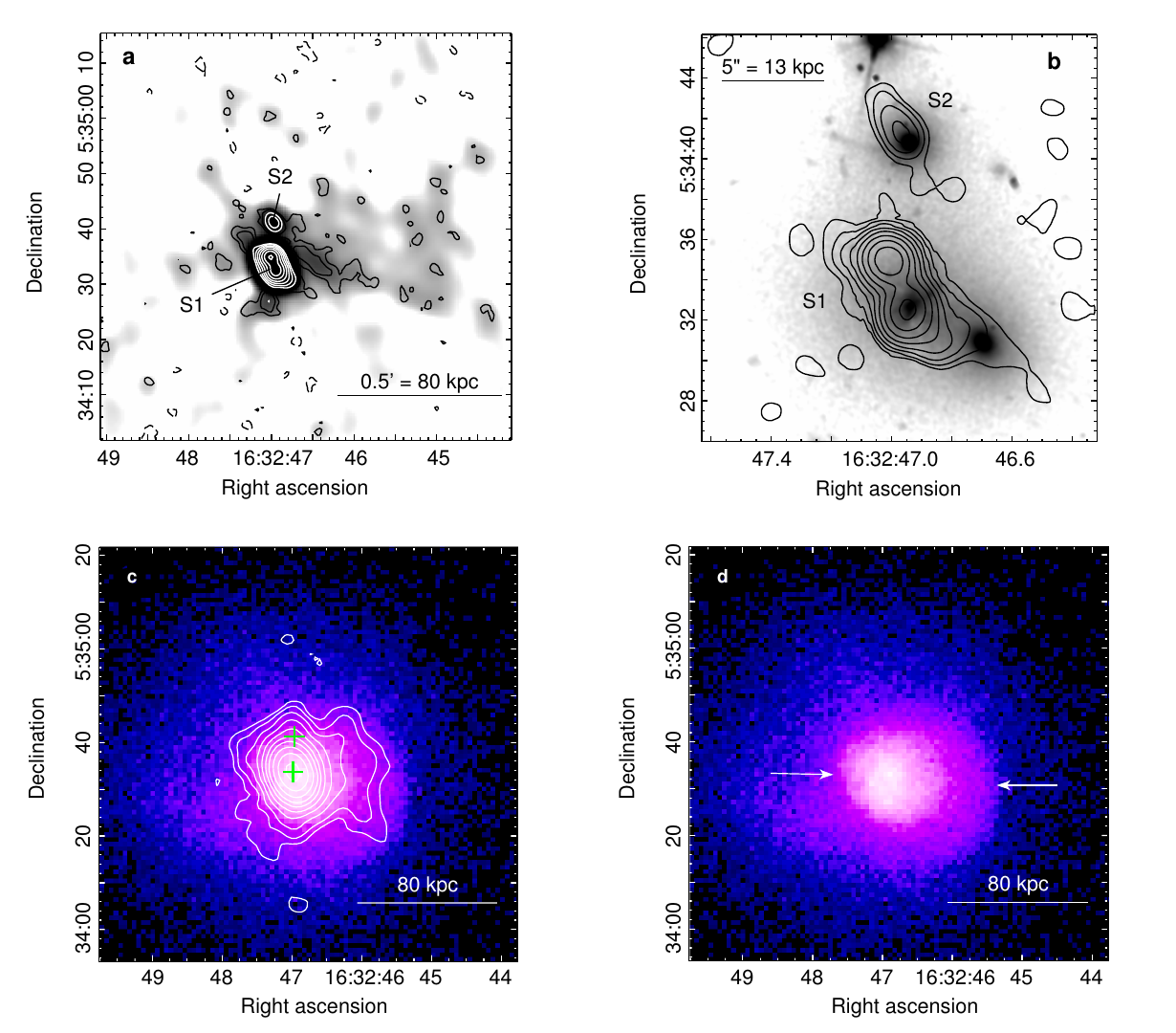}
\smallskip
\caption{ A\,2204. In all images, radio contours are $-1$ (dashed), 1, 2, 4, 8, 16,
... $\times 3\sigma$. (a) {\em VLA } 1.4 GHz contours (black and white) 
from the combined A+B array. The restoring beam is 
$2.0^{\prime\prime}\times1.9^{\prime\prime}$, in p.a. $77^{\circ}$ and the r.m.s. 
noise is $1\sigma=30 \, \mu$Jy beam$^{-1}$. The image is also shown 
in gray scale, smoothed with a Gaussian with a radius of 5 pixels
($2^{\prime\prime}$). S1 is the central radio galaxy and S2
is associated with a nearby galaxy (see panel(b) and Table 4). 
(b) {\em VLA}--A 1.4 GHz contours of S1 and S2, overlaid on the 
{\em HST} WFPC2 image (gray scale). The restoring beam is $1.4\times1.3$, 
in p.a. $45^{\circ}$ and r.m.s. noise is $1\sigma=25 \, \mu$Jy
beam$^{-1}$. No levels at $-3\sigma$ are
present in the portion of the image shown.
(c) {\em VLA}--B contours at 1.4 GHz of the
minihalo and discrete radio sources (green crosses), 
overlaid on the {\em Chandra} image in the 0.5-4 keV band.
The restoring beam 
is $6.0^{\prime\prime}\times5.0^{\prime\prime}$, in p.a., $17^{\circ}$
and the r.m.s. noise is $1\sigma=30 \, \mu$Jy beam$^{-1}$. 
No levels at $-3\sigma$ are present in the portion of the image shown.
(d) {\em Chandra} image, same as in (c). Arrows
indicate the position of two cold fronts.}
\label{fig:a2204}
\end{figure*}
%
%

%
%
\begin{figure*}
\centering
\includegraphics[scale=1.3]{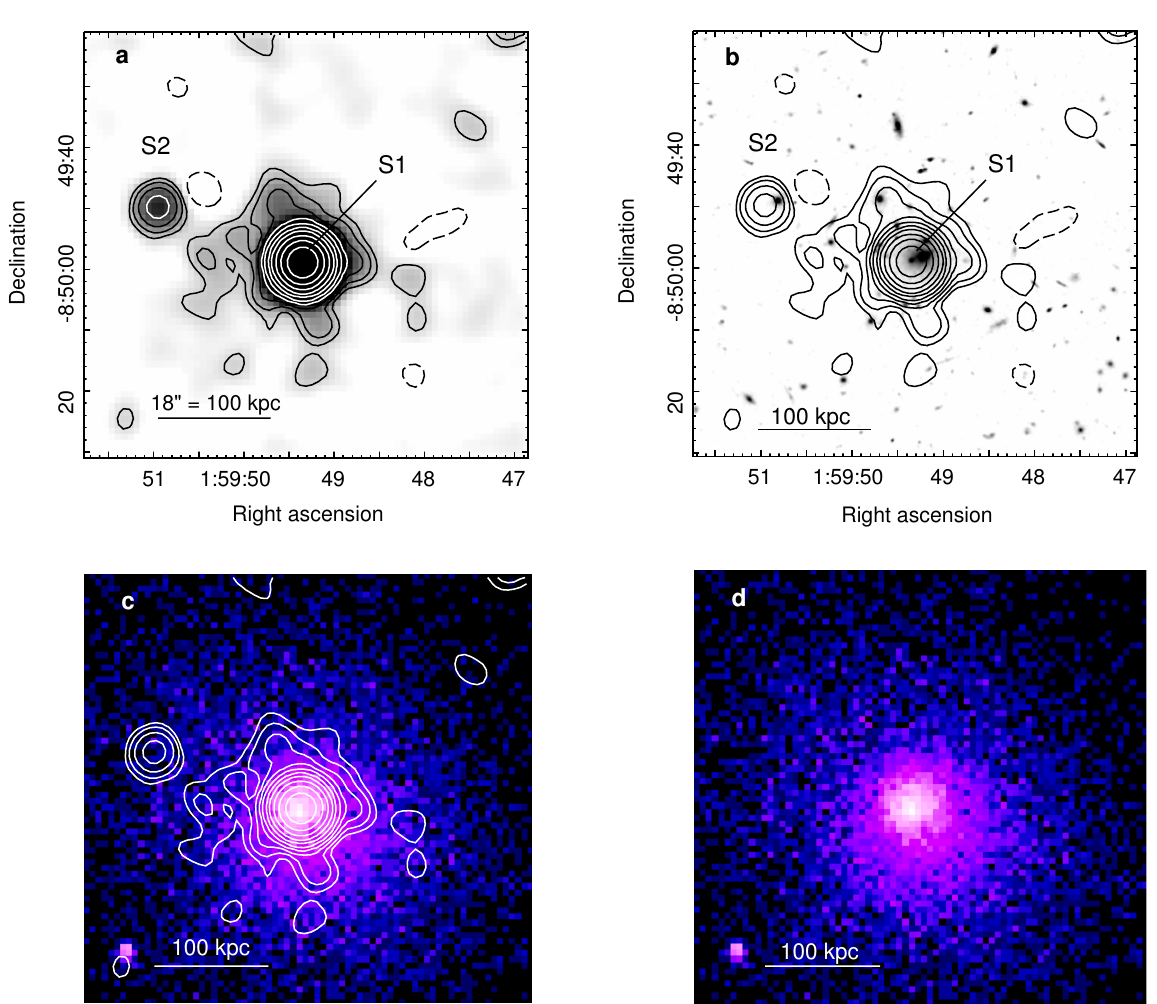}
\smallskip
\caption{ MACS\,J0159.8--0849. (a) {\em VLA} B-array contours at 1.4 GHz 
(black and white), restored with a $5^{\prime\prime}$ circular beam. 
The r.m.s. noise is $1\sigma=15\, \mu$Jy beam$^{-1}$ and 
contours are $-1$ (dashed), 1, 2, 4, 8, 16,... $\times 3\sigma$.
The image is also shown in gray scale, smoothed with a Gaussian with a 
radius of 3 pixels ($3^{\prime\prime}$). The unrsolved source S1 
is associated with the BCG and S2 is a possible cluster member 
(see panel (b) and Table 4). 
(b) 1.4 GHz contours, same as in (a), overlaid on the {\em HST} WFPC2
image (gray scale). (c) 1.4 GHz contours, same as in (a), overlaid on the 
{\em Chandra} image in the 0.5-4 keV band. The image has the same physical
size as panel (a). (d) {\em Chandra} image, same as in (c).}
\label{fig:macs0159}
\end{figure*}
%
%

%
%
\begin{figure*}
\centering
\includegraphics[scale=1.5]{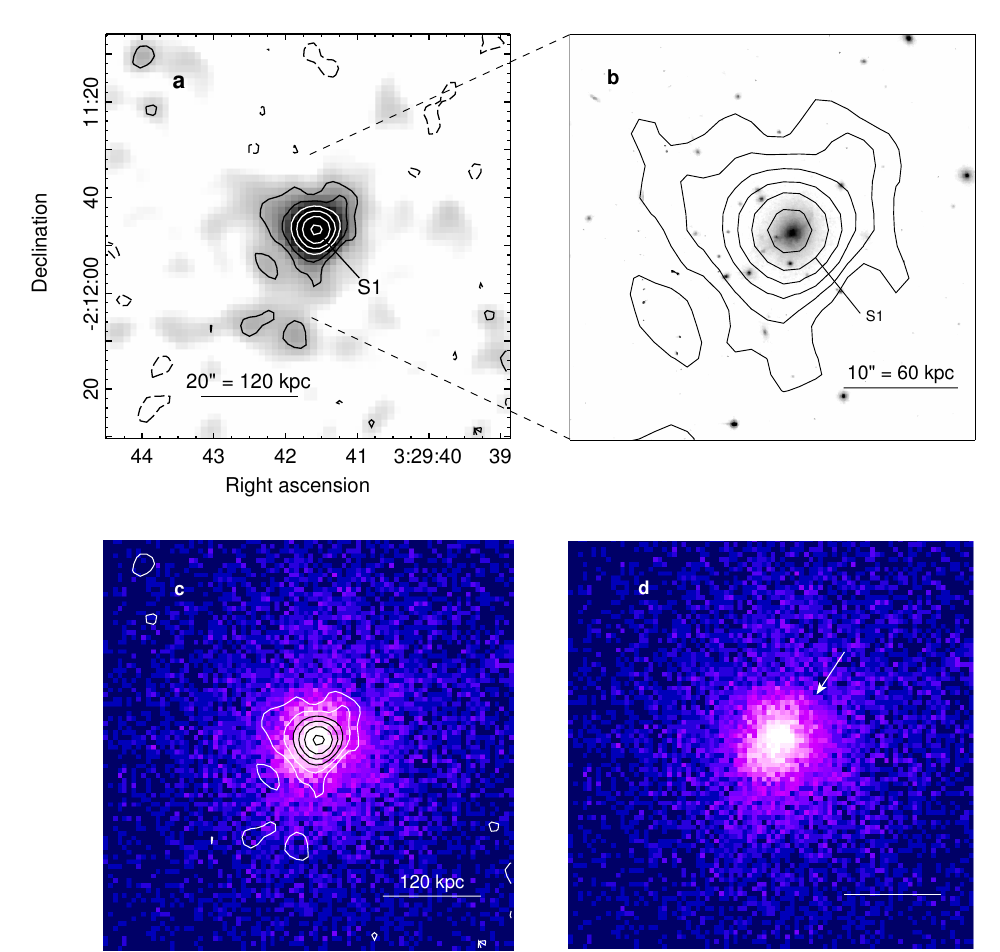}
\smallskip
\caption{ MACS\,J0329.6--0214. In all images, radio contours are 
$-1$ (dashed), 1, 2, 4, 8, 16,... $\times 3\sigma$.
(a) {\em VLA} 1.4 GHz contours (black
  and white) at the resolution of $4.7^{\prime\prime}\times4.5^{\prime\prime}$, in
p.a. $0^{\circ}$ (image credit: NRAO/{\em} VLA Archive Survey, (c) 2005-2007 AUI/NRAO). 
The r.m.s. noise is $1\sigma=40 \, \mu$Jy
beam$^{-1}$. The image is also shown 
in gray scale, smoothed with a Gaussian with a radius of 3 pixels
($4^{\prime\prime}$). The source S1 is
associated with the BCG (see panel (b) and Table 4).
(b) 1.4 GHz contours (same as in (a)), overlaid on the {\em HST} ACS
image (gray scale). (c) 1.4 GHz contours (black and white, 
same as in (a)), overlaid on the 
{\em Chandra} image in the 0.5-4 keV band. 
The image has the same physical scale as panel (a).
(d) 
{\em Chandra} image, same as in (c). The
arrow indicates a surface brightness edge at 
$\sim 10^{\prime\prime}$ from the center, which may be a cold front.}
\label{fig:macsj0329}
\end{figure*}
%
%

%
%
\begin{figure*}
\centering
\includegraphics[scale=1.7]{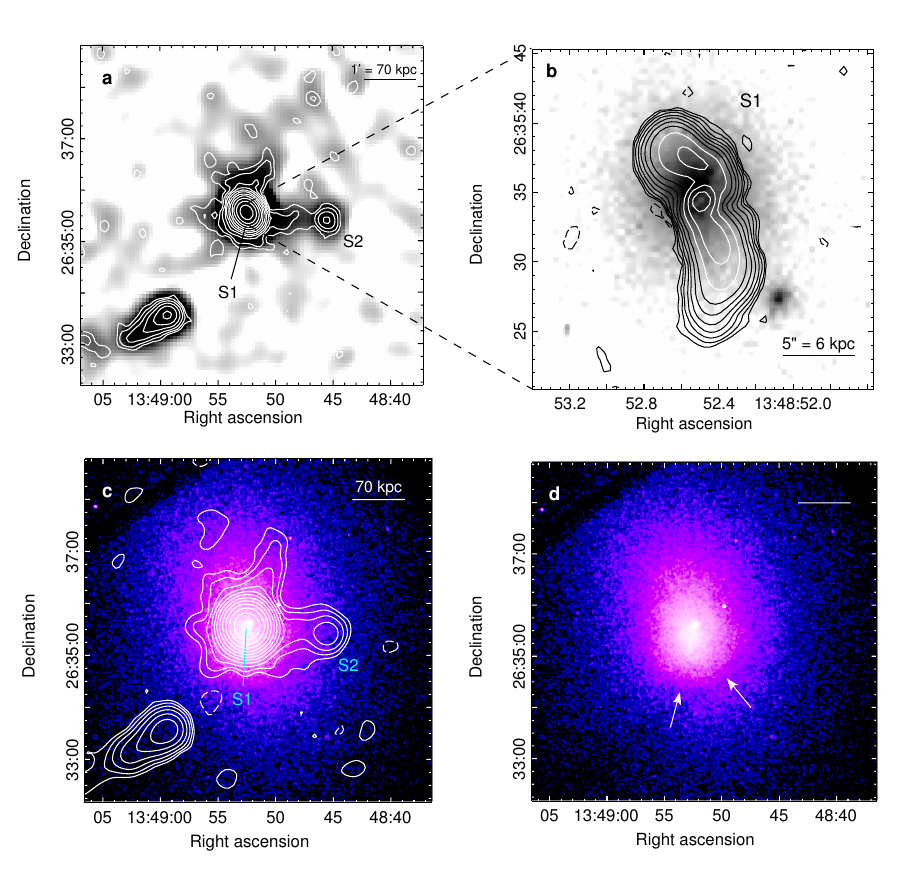}
\smallskip
\caption{A\,1795. In all images, radio contours are $-1$ (dashed), $1,
  2, 4, 8, 16, ... \times3\sigma$. (a) {\em VLA} C--array contours at 1.4
  GHz (white and black) at the resolution of
$18.7\times16.0$, in p.a. $37^{\circ}$. The r.m.s. noise is $1\sigma=150
\, \mu$Jy beam$^{-1}$. The image is also shown in gray scale, smoothed 
with a Gaussian with a radius of 3 pixels
($12^{\prime\prime}$). S1 is the cluster dominant radio galaxy
(4C\,26.42; see panel )b)), unresolved at this resolution, 
and S2 is a point source associated with a background galaxy (Table 4). 
(b) {\em VLA} A--array contours at 1.4 GHz of S1 (black and white), 
overlaid on the {\em HST} WFPC2 image (gray scale). The resolution is
$1.3\times1.2$, in p.a. $69^{\circ}$ and $1\sigma=40$
$\mu$Jy beam$^{-1}$. (c) {\em VLA} C--array contours at 1.4 GHz, 
overlaid on the 
{\em Chandra} image in the 0.5-4 keV band. The radio image has
been restored with a $27^{\prime\prime}$ circular beam. The
r.m.s. noise is $1\sigma=140 \, \mu$Jy beam$^{-1}$. (d) 
{\em Chandra} image, same as in (c). Arrows show the
position of a cold front.}
\label{fig:a1795}
\end{figure*}
%
%

%
%
\begin{figure*}
\centering
\includegraphics[scale=1.35]{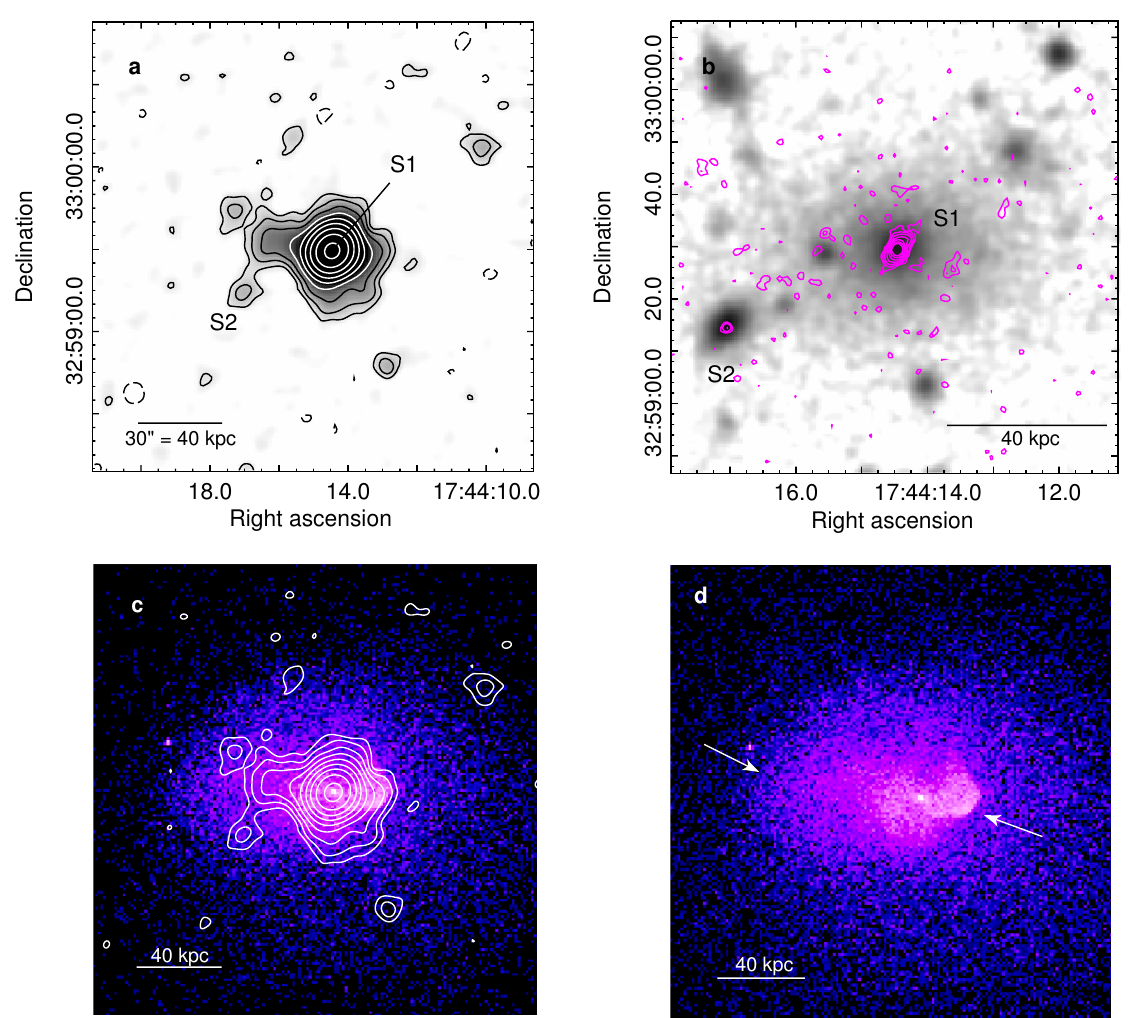}
\smallskip
\caption{ZwCl 1742.1+3306. (a) {\em VLA} DnA-configuration 
contours at 1.4 GHz (black and white). The restoring beam is 
$10.7\times9.7$, in p.a. $-40$. The r.m.s. noise is $1\sigma=70 \, \mu$Jy beam$^{-1}$ 
and contours are $-1$ (dashed), 1, 2, 4, 8, 16,... $\times 3\sigma$.
The image is also shown in gray scale, smoothed with a Gaussian with a 
radius of $3^{\prime\prime}$. The unresolved source S1 
is associated with the BCG and S2 is concident with
a nearby galaxy without redshift (see panel (b) and Table 4). (b) {\em VLA}--DnA
full-resolution contours at 1.3 GHz (magenta), overlaid on the optical 
POSS--2 image (gray scale). The restoring beam 
is $1.50^{\prime\prime}\times1.49^{\prime\prime}$,
in p.a., $48$ and $1\sigma=65 \, \mu$Jy beam$^{-1}$. Contours scale as
in (a). No levels at $-3\sigma$ are present
in the portion of the image shown. 
(c) 1.3 GHz contours (same as in (a)) overlaid on the {\em Chandra} image in 
the 0.5-4 keV band. The image has the same physical size as 
panel (a). (d) {\em Chandra} image, same as in (c).
Arrows indicate the position of two cold fronts.}
\label{fig:z1742}
\end{figure*}
%
%

%
%
\begin{figure*}
\centering
\includegraphics[scale=1.5]{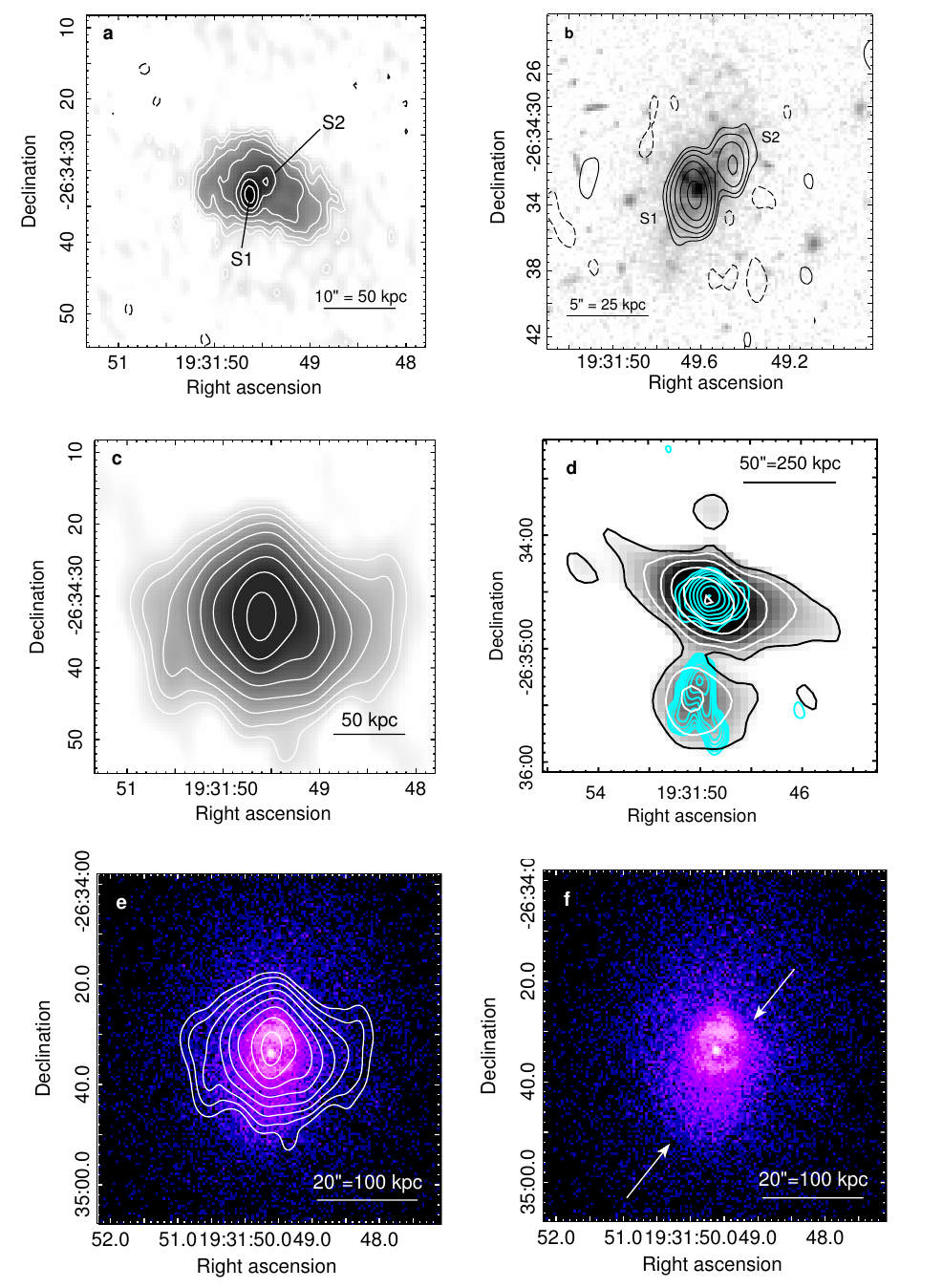}
\smallskip
\caption{MACS\,J1931.8--2634. In all radio images, contours are 
$-1$ (dashed), 1, 2, 4, 8, 16, ... $\times 3\sigma$, unless stated otherwise. (a) 
{\em VLA} A-configuration image (gray scale and black and white contours) 
at 1.4 GHz. The beam is $2.5\times1.2$, p.a. $3^{\circ}$ and $1\sigma=30 \, \mu$Jy beam$^{-1}$. S1 and S2 are the two 
central compact sources (see panel (b) and Table 4). 
(b) {\em VLA} A-configuration contours at 1.4 GHz 
from an image obtained using only baselines longer than 20 $k\lambda$, 
overlaid on the {\em HST} WFPC2 image (gray scale). The beam is
$2.4\times1.0$, p.a. $2^{\circ}$ and $1\sigma=60 \, \mu$Jy beam$^{-1}$. 
(c) {\em VLA} B-configuration image at 1.4 GHz (contours and
gray scale). The beam 
is $8.4^{\prime\prime}\times3.8^{\prime\prime}$, in p.a.,
$0^{\circ}$ and $1\sigma=40 \, \mu$Jy beam$^{-1}$. No
levels at $-3\sigma$ are
present in the portion of the image shown. (d) {\em GMRT} 150 MHz
image from the TGSS (white contours and gray scale), with overlaid
the {\em VLA} 1.4 GHz contours (cyan) from panel (c). The beam 
is $25^{\prime\prime}\times15^{\prime\prime}$, in p.a.,
$30^{\circ}$ and $1\sigma=25$ mJy beam$^{-1}$. 
White contours scale by a factor of 2 starting from 
100 mJy beam$^{-1}$. No levels at $-100$ mJy beam$^{-1}$ are
present in the portion of the image shown.
(e) 1.4 GHz contours (same as in panel (c)), overlaid on the 
{\em Chandra} image in the 0.5-4 keV band. (e) {\em Chandra} image,
same as in (e). Arrows show the position of two possible cold fronts. 
Two small cavities are visible to the east and west of the 
central AGN (see Ehlert et al. 2011).}
\label{fig:macs1931}
\end{figure*}
%
%

\subsection{ZwCl\,3146}\label{sec:3146}

ZwCl\,3146 is hot ($kT\sim7.5$ keV; Table 5) and relaxed cluster 
at $z=0.29$. It
exhibits a pronounced cool core with one of the strongest cooling
rate known \citep[$\sim 1600$ $M_{\odot}$ per year,][]{2013EGUGA..15.7425K}.
Multiple cold fronts have been detected by {\em Chandra} in its core 
\citep{2002astro.ph..7165F}; two of them are visible at small radii 
($\sim 3^{\prime\prime}$ and $8^{\prime\prime}$, corresponding to
$\sim 13$ kpc and $\sim 35$ kpc), while the third front is located 
at $\sim 35^{\prime\prime}$ ($\sim 150$ kpc) from the cluster center
(see also Fig.~\ref{fig:zw3146}(d)).

No pointed {\em VLA} observations at 1.4 GHz exist for ZwCl\,3146. 
Inspection of the 1.4 GHz image from the NVSS\footnote{NRAO {\em VLA} Sky Survery,
\cite{1998AJ....115.1693C}.} indicates that the cluster hosts 
a central radio source, whose structure is unresolved at the $45^{\prime\prime}$
resolution of the image. A Gaussian fit to the source gives an integrated 
flux density of $8.7\pm0.5$ mJy.

We used {\em VLA} observations at 4.9 GHz and 8.5 GHz to image the 
radio source at higher resolutions (Table 2). A central point source,
embedded in faint, extended emission, is detected in the 4.9 GHz 
$\sim 5^{\prime\prime}$-resolution image (Fig.~\ref{fig:zw3146}(a)). 
The diffuse, amorphous feature, which we classify as a minihalo, has a
radius of $\sim 90$ kpc. A second point source (S2) is detected 
south-west of S1. The $2^{\prime\prime}$-resolution image at 8.5 GHz is 
shown in Fig.~\ref{fig:zw3146}(b), overlaid on the optical
{\em  HST} image. Both point 
sources have optical counterparts: S1 is identified with the BCG 
and S2 is associated with a background galaxy 
with a photometric redshift $z_{\rm  phot}=0.34$ from the SDSS (Table 4).
No extended emission connected to these sources, such as jets or lobes, 
is detected. The flux densities of S1 and 
S2 are summarized in Table 4. Their spectral indices 
are $\alpha=0.66\pm0.14$ (S1) and $\alpha=-0.33\pm0.16$ (S2). 
After subtraction of the flux densities of S1 and S2 from the
total emission in Fig.~\ref{fig:zw3146}(a), a flux of $1.7\pm0.2$ mJy 
is found for the minihalo at 4.9 GHz. Assuming a power-law spectrum 
extrapolated from the 4.9--8.5 GHz frequencies, the expected flux densities 
of S1 and S2 at 1.4 GHz are 3.3 mJy and 0.2 mJy, respectively.
Subtraction of these values from the total flux density in the 
NVSS image provides an estimate of the flux density of the 
minihalo at 1.4 GHz of $\sim 5.2$ mJy. The corresponding radio power is 
$\sim 1.4\times10^{24}$ W Hz$^{-1}$.
 and the estimated 1.4--4.9 GHz spectral index of the minihalo is $\alpha \sim 1$.

The minihalo is overlaid on the {\em Chandra} image in 
Fig.~\ref{fig:zw3146}(c), also shown in panel (d) 
with arrows marking the position of the cold fronts.
The minihalo emission extends mostly in the northern and SE sectors of the core.
No clear radio features are visible at the position of the cold fronts: 
the innermost front lies inside the minihalo, near the central 
point source; the outer fronts are at larger radii 
than the minihalo.

Because of their steep radio spectra ($\alpha \sim 1.2-1.3$) 
and low surface brightness, minihalos are usually best detected 
at frequencies around 1 GHz or lower. There are very few detections 
at higher frequencies, e.g., 2A\,0335+096 \citep[][Giacintucci et al.
in prep.]{1995ApJ...451..125S}, RXCJ\,1720.1+26 (Giacintucci et al. in preparation) 
and RX\,J1532.9+3021 (this paper). It is therefore noteworthy that this
minihalo has been discovered at 4.9 GHz.

\subsection{A\,2204}

Abell 2204 is a $\sim 9$ keV cluster (Table 5) at $z=0.152$, 
regular and relaxed on large scales, but disturbed in its cool core, 
where metallicity substructures and multiple X-ray cavities have been
revealed by {\em Chandra} \citep{2005MNRAS.356.1022S,2009MNRAS.393...71S}.
{\em Chandra}
also showed a clear spiral structure in the core surface brightness
distribution, defined by two 
cold fronts at $\sim 28$ kpc and $\sim 55$ kpc from the center 
\citep[][see also Fig.~\ref{fig:a2204}(d)]{2005MNRAS.356.1022S,2010ApJ...717..908Z}.
A small radio source 
is associated with the BCG \citep{2009MNRAS.393...71S}. 
At arcsecond resolution, 
it consists of a compact, flat-spectrum core and two
weak, extended components to the north and south.
Hints of further extended emission to the east and west of the central
radio galaxy were reported by \cite{2005MNRAS.356.1022S,2009MNRAS.393...71S},
who suggested the possible existence of a larger-scale minihalo.

We reprocessed the {\em VLA} data at 1.4 GHz analyzed by \cite{2009MNRAS.393...71S}.
Our radio images are presented in Fig.~\ref{fig:a2204}. In the images 
from the A and combined A+B configurations (panels (a) and (b)), we detect the central 
cluster radio galaxy (S1) and a second unresolved source (S2), 
associated with a galaxy without redshift information. Their morphology and
flux densities (Table 4) are consistent with those reported by 
\cite{2009MNRAS.393...71S}. The smoothed gray-scale image in panel 
(a) shows the diffuse emission west of S1, first noticed by 
\cite{2005MNRAS.356.1022S}. 
To investigate this component, we produced an image from the  
B array applying natural weighting. The image is overlaid on the
{\em Chandra} image in Fig.~\ref{fig:a2204}(c). 
Diffuse radio emission is clearly present in the area around S1 and S2 (crosses),
confirming the initial suggestion of \cite{2005MNRAS.356.1022S}
of a centrally-located minihalo. The source is $\sim 50$ kpc in radius
and has a total flux density of $8.6\pm0.9$ mJy (after subtraction of S1 and S2), which
corresponds to a radio power of $(5.4\pm0.5)\times10^{23}$ W Hz$^{-1}$.

In Fig.~\ref{fig:a2204}(d), arrows mark the position of the
two cold fronts that define the prominent spiral pattern in the {\em Chandra} image. 
The minihalo emission appears to be contained within the region bounded 
by the fronts, as observed in other minihalo clusters with similar 
spiral-like patterns in their cool cores \citep{2008ApJ...675L...9M,2013ApJ...762...78Z}.

\section{Candidate minihalos}

In 6 of the clusters listed in Table 1, we found 
diffuse radio emission around the dominant radio galaxies, 
but whose classification as a minihalo is uncertain and requires 
further observations. The radio images of these new candidate
minihalos are discussed here.

\subsection{MACS\,J0159.8--0849}\label{sec:0159}

MACS\,J0159.8--0849 is a relatively high-redshift 
($z=0.405$), hot ($\sim 9$ keV; Table 5) cluster with regular X-ray 
morphology \citep[e.g.,][]{2008ApJS..174..117M}. The X-ray surface 
brightness profile peaks at the BCG and the gas temperature 
declines toward the center \citep[e.g.,][]{2009ApJS..182...12C},
as typically observed in other cool-core clusters.

We imaged the central region of the cluster 
at 1.4 GHz using {\em VLA} observations in B configuration (Table 2). 
Our image is shown in Fig.~\ref{fig:macs0159}, where it 
is also overlaid on the {\em HST} optical and {\em Chandra} X-ray images. 
We detected a point source (S1) at the BCG 
and a second unresolved source $\sim 0.4^{\prime}$ to the NE,
associated with a possible cluster member galaxy with $z_{\rm phot}=0.44$ (Table 4).
We also detect a $\sim 90$ kpc-radius, diffuse source around S1, which 
is probably a minihalo. To resolve S1 and investigate its connection 
with the larger-scale extended emission, we checked the {\em VLA} 
archive for higher-resolution observations. There is a pointed 
observation at 8.5 GHz in A configuration, which we used to obtain an 
image of the BCG at sub--arcsecond resolution (not shown here). We found a 
bright point source coincident with the peak of S1 in the 1.4 GHz image. 
No other radio emission is detected at the sensitivity of the 8.5 GHz image 
($1\sigma=18$ $\mu$Jy beam$^{-1}$). 

We measured the flux density of S1 at 1.4 GHz on an image obtained 
excluding the innermost $15$ k$\lambda$-region of the $u-v$ plane, which 
is sensitive to the larger-scale emission, and found $36.7\pm1.8$ mJy. Comparison 
between the flux densities at 1.4 GHz and 8.5 GHz ($94.8\pm4.8$ mJy) gives $\alpha=-0.55\pm0.04$, 
suggesting that S1 is likely dominated by a bright core with an inverted spectrum.
To estimate the flux density of the candidate minihalo at 1.4 GHz, 
we subtracted the flux density of S1 and S2 from the total emission
measured in Fig.~\ref{fig:macs0159}(a), 37.8 mJy. We found 
$2.4\pm0.2$ mJy in diffuse emission, which corresponds to a radio power of 
$(1.40\pm0.14)\times10^{24}$ W Hz$^{-1}$.

The comparison between the radio and X-ray {\em Chandra} images
is shown in Fig.~\ref{fig:macs0159}(c). The diffuse radio emission 
permeates most of the core region, with a similar roundish distribution. 
No obvious features are seen in the X-ray surface brightness.  

The confirmation of the minihalo nature of the central extended emission 
requires further observations. In particular, higher-resolution images are 
needed to determine the morphology of the central radio galaxy and whether 
it has any connection to the surrounding
emission. MACS\,J0159.8--0849 has been recently observed with the 
GMRT at 325 MHz as part of a cluster survey (Macario et al.
in preparation). Diffuse emission is visible in the 325 MHz image 
on a similar spatial scale. However, the angular resolution of 
the GMRT at this frequency ($\sim 10^{\prime\prime}$) does not allow to
separate the central source from the extended emission.

\subsection{MACS\,J0329.6--0211}\label{sec:0329}

MACS\,J0329.6--0211 is the most distant ($z=0.45$) among the clusters 
studied here. It is a relaxed system with a global temperature of
$\sim 6$ keV (Table 5) and a bright, low-entropy cool core \citep{2009ApJS..182...12C}.

Only one pointed observation at 1.4 GHz is available in the 
{\em VLA} archive (Table 2). The observation is in B configuration and the
resulting image is presented in Fig.~\ref{fig:macsj0329}(a).
A bright point source (S1) is coincident with the position of 
the BCG, as shown by the overlay of the radio emission on 
the optical {\em HST} image (b).
The compact source appears surrounded by a heart-shaped 
diffuse structure, possibly a minihalo, with a radius of
 $\sim 70$ kpc ($+3\sigma$ isocontour). A slighty smoothed 
image at 1.4 GHz is shown as gray scale in Fig.~\ref{fig:macsj0329}(a)
and includes emission below the $3\sigma$ level. 
The image suggests that the candidate minihalo extends further out, 
possibly reaching a distance of $\sim 150$ kpc from the center, 
if we include the positive residuals visible to the 
south-east. 

For S1, we measure a flux density of $3.8\pm0.2$ mJy and
a radio power of $(2.8\pm0.1)\times 10^{24}$ W Hz$^{-1}$ on 
an image obtained cutting the innermost $15$ k$\lambda$-region of the $u-v$ plane
to remove the contribution of the sourrounding extended emission.
The candidate minihalo has $3.8\pm0.4$ mJy and 
$P_{\rm 1.4 \, GHz}=(2.8\pm0.3) \times 10^{24}$ W Hz$^{-1}$.
 
An overlay of the radio contours on the {\em Chandra} image in
Fig.~\ref{fig:macsj0329}(c) indicates that the diffuse radio emission 
is centrally located and fills a large portion of the cluster core, as 
seen in other minihalos clusters. The cluster appears very relaxed 
on large scales. In the core, a surface brightness edge, possibly 
a cold front, is visible at $\sim 10^{\prime\prime}$ from the center, 
indicated by the arrow in panel (d).

Deeper radio observations are needed to determine the total 
extent of the putative minihalo. In addition, 
as in the case of MACS\,J0159.8--0849, radio 
data at higher resolution are necessary to image the radio 
emission associated with the BCG and investigate its possible
connection to the diffuse component.

\subsection{A\,1795}

A\,1795 is one of the first clusters in which a sloshing cold front has been 
detected with {\em Chandra} \citep{2001ApJ...562L.153M}.
This nearby ($z=0.06$) cool-core cluster is considered one of the most relaxed 
clusters in X-rays, showing a very regular X-ray morphology \citep[e.g.,][]{2002MNRAS.331..635E}. Its cD harbors an FR-I radio galaxy (4C\,26.42) with a double 
Z-shaped morphology and a size of $\sim 12^{\prime\prime}$ \citep[$\sim 14$ kpc;][]{1993AJ....105..778G}.

Our {\em VLA} images at 1.4 GHz are presented in Fig.~\ref{fig:a1795}. 
Panel (a) shows the C-array image ($\sim 19^{\prime\prime}$ 
resolution), in which 4C\,26.42, labelled S1, is unresolved. A second point source (S2) 
is detected $\sim 1^{\prime}.6$ west of S1. This source coincides with a 
background galaxy with $z_{\rm  phot}=0.57$ (Table 4). Fig.~\ref{fig:a1795}(b) 
shows an arcsecond-resolution image of S1 \citep[in good agreement with][]{1993AJ....105..778G}, overlaid on the optical {\em HST} image of the BCG.  
Fig.~\ref{fig:a1795}(a) indicates that larger-scale diffuse emission 
may be present around S1 and in the region between S1 and S2, 
suggesting the existence of a central minihalo. 
The brightness distribution of the candidate minihalo appears unusually
filamentary, with two prominent radial extensions pointing northwest and west. 
The smoothed gray-scale image suggests that the minihalo 
could extend to a radius of $\sim 100$ kpc, as also 
shown by an image with a slightly lower resolution,
overlaid on the {\em Chandra} image in Fig.~\ref{fig:a1795}(c). 

To measure the flux density of the candidate minihalo,
we subtracted the contribution of S1 and S2 from the total 
flux density in our low resolution image. For S1, we used the flux density 
measured on the FIRST\footnote{Faint Images of the Radio Sky at Twenty-cm 
\citep{1995ApJ...450..559B}.} image ($917\pm46$ mJy; not shown here), 
where the radio galaxy is slightly extended, with its major axis oriented in NE-SW, 
as seen at higher resolution. We note that the FIRST value is 
between the $890\pm45$ mJy measured on the A-array image 
and $960\pm48$ mJy given by a Gaussian fit 
to the central point source in the C-array image.
For S2, we measured $5.1\pm0.3$ mJy. The resulting flux of the minihalo
candidate is $85\pm5$ mJy, corresponding to a
radio power at 1.4 GHz of $(7.9\pm0.5)\times10^{23}$ W Hz$^{-1}$.

The prominent cold front discovered by \cite{2001ApJ...562L.153M}
in the 
{\em Chandra} image is located  $\sim 70^{\prime\prime}$ south of 
the center (arrows in
Fig.~\ref{fig:a1795}(d)). The radio/X-ray comparison in panel (c) 
suggests that the cold front may be confining the southern emission 
of the minihalo.

Deeper radio observations are necessary to confirm the presence of a central 
minihalo in this cluster and determine its total size and possible 
association with the cold front.

\subsection{ZwCl\,1742.1+3306}\label{sec:1742}

ZwCl\,1742.1+3306 is a nearby ($z=0.076$), cool-core cluster
with a global temperature of $\sim 5$ keV (Table 5). A recent {\em Chandra} 
detection of two well-defined cold fronts at $\sim 20$ kpc west and 
$\sim 80$ kpc east of the BCG indicates that the cool core is 
sloshing \citep[][see also Fig.~\ref{fig:z1742}(d)]{2013A&A...555A..93E}. 
Two possible small cavities are located a few tens of kpc north and south 
of the BCG \citep{2013A&A...555A..93E}. 

We imaged the cluster center at 1.4 GHz using the DnA-configuration 
observation in Table 2. The angular resolution of the full array is 
$1^{\prime\prime}$. We also obtained images with slighly lower 
angular resolutions by weighting down long baseline data points. 
Our lowest-resolution image ($\sim 10^{\prime\prime}$) is shown 
in Fig.~\ref{fig:z1742}(a). Extended emission is 
present at the cluster center, around the central radio galaxy 
S1. At arcsec resolution, the source S1 is dominated by a bright 
compact component (Fig.~\ref{fig:z1742}(b)), whose peak is 
coincident with the central AGN, detected as a point source in 
X-rays \citep[][see also Fig.~\ref{fig:z1742}(d)]{2013A&A...555A..93E}. 
Two weak, extended components are located NW and SW of the center, 
pointing to, but not reaching, the two X-ray cavities 
(Fig.~\ref{fig:z1742}(d)). A point source (S2) is detected at 
the position of a galaxy without redshift information. 

\cite{1991ApJS...75.1011G} report a flux density of 74 mJy at 4.9 GHz 
for the central radio galaxy. We measure 69 mJy at 1.4 GHz, which
implies a flat (possibly inverted) spectrum ($\alpha=-0.1$). 
This is consistent with the core-dominated morphology of S1
in Fig.~\ref{fig:z1742}(b). After the subtraction of the flux densities of 
S1 and S2 (Table 4) from the total emission in Fig.~\ref{fig:z1742}(a), 
we estimate a residual flux density of $13.8\pm0.8$ mJy in diffuse emission, 
corresponding to a luminosity of $(2.0\pm0.1)\times10^{23}$ W Hz$^{-1}$. 

A bright, unresolved source is detected at the cluster center by the 
VLSSR\footnote{{\em VLA} Low-Frequency Sky Survey Redux \citep{2012RaSc...47.0K04L}} 
at 74 MHz and WENNS\footnote{The Westerbork Northern Sky Survey \citep{1997A&AS..124..259R}} 
at 325 MHz, with flux densities of 2.4 Jy\footnote{this value is in the \cite{2012MNRAS.423L..30S} flux density scale.}
and 332 mJy, respectively, giving $\alpha=1.3$ for the whole emission (S1+diffuse
component). Although higher-resolution images at
these frequencies are needed to separate the large-scale emission from 
the central source S1, this spectral index suggests that the extended 
emission seen in Fig.~\ref{fig:z1742}(b) has a steep spectrum.

Fig.~\ref{fig:z1742}(c) shows the {\em Chandra} image with an overlay 
of the 1.4 GHz contours. Within the central $r\sim30$ kpc, the diffuse radio 
emission is relatively round and covers the brightest X-ray emission in 
the core. On larger scales, the radio emission extends toward
east, out to a radius of $\sim$40 kpc, following an elongation of the X-ray 
emission toward the eastern cold front (Fig.~\ref{fig:z1742}(d)).
Overall, the radio emission appears contained within the 
sloshing region defined by the cold fronts, though at this brightness
level it does not reach the eastern front. 

As we will discuss in \S~\ref{sec:1931}, the nature of the 
extended emission detected in this cluster is unclear. 
Further radio observations are necessary to investigate the
relationship between the central radio galaxy and the larger-scale
component.

\subsection{MACS\,J1931.8--2634}\label{sec:1931}

MACS\,J1931.8--2634 ($z=0.35$) is a relaxed and massive cluster
that harbors one of the most X-ray luminous cool cores known, with a bolometric 
luminosity of $\sim 1\times10^{45}$ erg s$^{-1}$ and a cooling time $<1$ Gyr 
in its innermost $r\sim 50$ kpc region \citep{2011MNRAS.411.1641E}.

A combined X-ray, optical and radio study of this system 
has been presented by \cite{2011MNRAS.411.1641E}, showing that the 
cluster core has been significantly disturbed by both 
AGN activity and sloshing of the low-entropy gas along 
a roughly N-S direction, as indicated by a prominent
spiral structure in the temperature map. Two possible 
X-ray cavities are located to the east and west of the central 
powerful AGN, which is a bright X-ray source \cite[][see also
Fig.~\ref{fig:macs1931}(f)]{2011MNRAS.411.1641E}.
Using the {\em VLA}, \cite{2011MNRAS.411.1641E} imaged the cluster 
core at 1.4 GHz and found a central compact source, coincident with
the BCG, and extended, amorphous emission on a larger scale.

We reprocessed these {\em VLA} observations (Table 2). Our 
images are shown in Fig.~\ref{fig:macs1931}. As previously found,
the emission in the cluster core is composed by a bright, unresolved 
source (S1), wholly enclosed in a much fainter, diffuse structure
$\sim 100$ kpc long in the E-W direction and $\sim 60$ kpc in the perpendicular 
direction (Fig.~\ref{fig:macs1931}(a)). No distinct jets or lobes are seen in 
the extended emission.
A second, compact component (S2) is detected $\sim 3^{\prime\prime}$ 
from S1. To better separate S1 and S2 from the surrounding extended 
emission, we produced an image from the A-configuration data using 
only the baselines longer than $20$ k$\lambda$.
In Fig.~\ref{fig:macs1931}(b) we show the resulting image overlaid on the optical {\em HST} 
image of the BCG. The peak of S1 is coincident with the central 
AGN source. S2 has no clear association with point sources 
on the {\em HST} image, but is cospatial 
with a bright H$\alpha$ and blue light 
filament to the north-west of the BCG, which also coincides with a 
bright, cool and metal-rich ridge of X-ray emission \citep{2011MNRAS.411.1641E}.

The 1.4 GHz image from the B configuration is presented in 
Fig.~\ref{fig:macs1931}(c). The source is more extended 
that imaged at higher resolution and reaches a size of 
$R_{\rm max}\sim 100$ kpc and $R_{\rm min} \sim 90$ kpc, 
almost twice 
the size seen in the A-configuration image in panel (a). The source 
is also detected with the {\em VLA} at 327 MHz on 
a similar spatial scale (image not shown here), but it is only 
marginally resolved due to the relatively low angular resolution 
(Table 2). Inspection of a 150 MHz image from the 
TGSS\footnote{TIFR GMRT Sky Survey (Sirothia et al. in preparation),
http://tgss.ncra.tifr.res.in/150MHz/tgss.html} 
reveals that the source is even more extended. The 150 MHz image is 
reported in panel (d) with the 1.4 GHz contours overlaid 
for comparison. The emission seen at low frequency appears 
slightly tilted with respect to the E-W emission at 1.4 GHz,
and reaches a large size of $R_{\rm max}\sim 300$ kpc along 
its NE-SW major axis. The marginally extended source to the south is resolved 
at 1.4 GHz into a narrow-angle tail (NAT) associated with a bright 
galaxy $\sim 40^{\prime\prime}$ (projected) from the 
BGC \citep[see also][]{2011MNRAS.411.1641E}.
 
The flux densities at 1.4 GHz of S1 and S2 are $11.6\pm0.6$ mJy and
$2.5\pm0.1$ mJy, respectively, as measured on the image in 
Fig.~\ref{fig:macs1931}(b). A total of $62\pm3$ mJy is found in the 
low-resolution, B-array image in Fig.~\ref{fig:macs1931}(c). 
After subtraction of the contributions of S1 and S2, $48\pm3$ mJy 
can be attributed to the outer diffuse emission, which is therefore the 
dominant component in the system with $\sim 80$ \% of the total flux density. 
Its radio power is $(2.0\pm0.1)\times10^{25}$ W Hz$^{-1}$. Due to the lower 
angular resolution of the 327 MHz and 150 MHz images, we are not able to 
separate the individual components and can only obtain a flux density for 
the whole system. 
We measure $669\pm33$ mJy at 327 MHz and $6.15\pm1.50$ Jy at 150 MHz.
The inferred spectral index between 150 MHz and 1.4 GHz is very steep, 
$\alpha=2.1\pm0.1$. 

In Fig.~\ref{fig:macs1931}(e), we overlay the 1.4 GHz contours on the 
{\em Chandra} image. The diffuse emission covers the entire core 
region and is far more extended than the region occupied by the 
possible X-ray cavities, whose association with the radio emission is 
therefore unclear. On the other hand, the axis connecting the 
cavities and major axis of the radio emission are oriented in a 
similar direction, suggesting a connection between 
the radio and X-ray features. Furthermore, as noticed by \cite{2011MNRAS.411.1641E},
the outer boundaries of the putative cavities are 
not well defined. It is therefore possible that the cavities 
are larger, or additional, older cavities are present further 
away from the center.

Due to the ultra-steep radio spectrum of the radio emission, 
its highly elongated and much wider structure at low frequency, 
and its unclear link with the X-ray cavities, 
we consider the classification of this source as a minihalo very
uncertain, as discussed in more detail in \S \ref{sec:1931}. 

\section{Discussion}
\label{sec:disc}

Our current knowledge of radio minihalos in cluster cores has been based 
on only 12 reported objects (Table 5), which is insufficient to investigate 
their origin and answer the question why they are so rare.
It is therefore crucial to look for more minihalos in a systematic way.
We have selected a large sample of X-ray luminous clusters with available 
high-quality radio data to search for new possible minihalos.
Our goal is to investigate the radio properties of these clusters 
in relation their global and core properties 
(Giacintucci et al. in preparation). In this paper, we have presented 
9 clusters from the sample, in which we found clear indication of a 
centrally-located diffuse radio emission around the radio active BCG
in their {\em VLA} images. Below we give a brief summary of our findings.

\begin{enumerate}

\item  Based on the available radio information at both high 
and low angular resolutions, we have found central diffuse 
emission in A\,478, RX\,J1532.9+3021, ZwCl\,3146 and A\,2204
that is consistent with a minihalo. In these clusters, we were able 
to image the central 
radio galaxy on a scale $r\lesssim 10$ kpc, that is considerably 
smaller than the extent of the surrounding minihalo ($R_{\rm MH}>50$ kpc), 
and found no connection (for instance, in the form of jets or lobes) 
between the radio galaxy and the outer extended emission.

\item  In MACS\,J0159.8--0849 and MACS\,J0329.6--0214, 
the two most distant systems among the clusters presented here, 
the existing data do not allow to map the emission 
from the central radio galaxy on scales $r\lesssim 100$ kpc. 
Thus, we cannot rule out a direct connection between radio galaxy 
and larger scale diffuse emission, which we therefore classify as 
candidate minihalos.

\item We found a candidate minihalo in A\,1795, where hints of a very faint, 
diffuse source of a much larger 
linear size than the central radio galaxy 4C\,26.42 can be seen 
in the 1.4 GHz image. 

\item We detected extended emission enshrouding the dominant radio galaxy in 
ZwCl\,1742.1+3306 and MACS\,J1931.8--2634, however the interpretation of 
this emission as a minihalo is uncertain and requires further radio observations. 
These sources will be discussed in more detail in \S \ref{sec:1931}.

\end{enumerate}

The new minihalos and candidates and the previously known minihalos 
reported in Table 5 are all well detected. Even for the least significant detection 
-- the candidate minihalo in MACS\,J0329.6--0214 -- the brightest part of the minihalo 
outside of the central point source is imaged at $\sim 7\sigma$ with respect to the image 
noise (per beam) and the best cases are imaged at more than $50\sigma$.

\subsection{Radio properties of the BCG in minihalo clusters}

\begin{figure}
\centering
\includegraphics[scale=0.8]{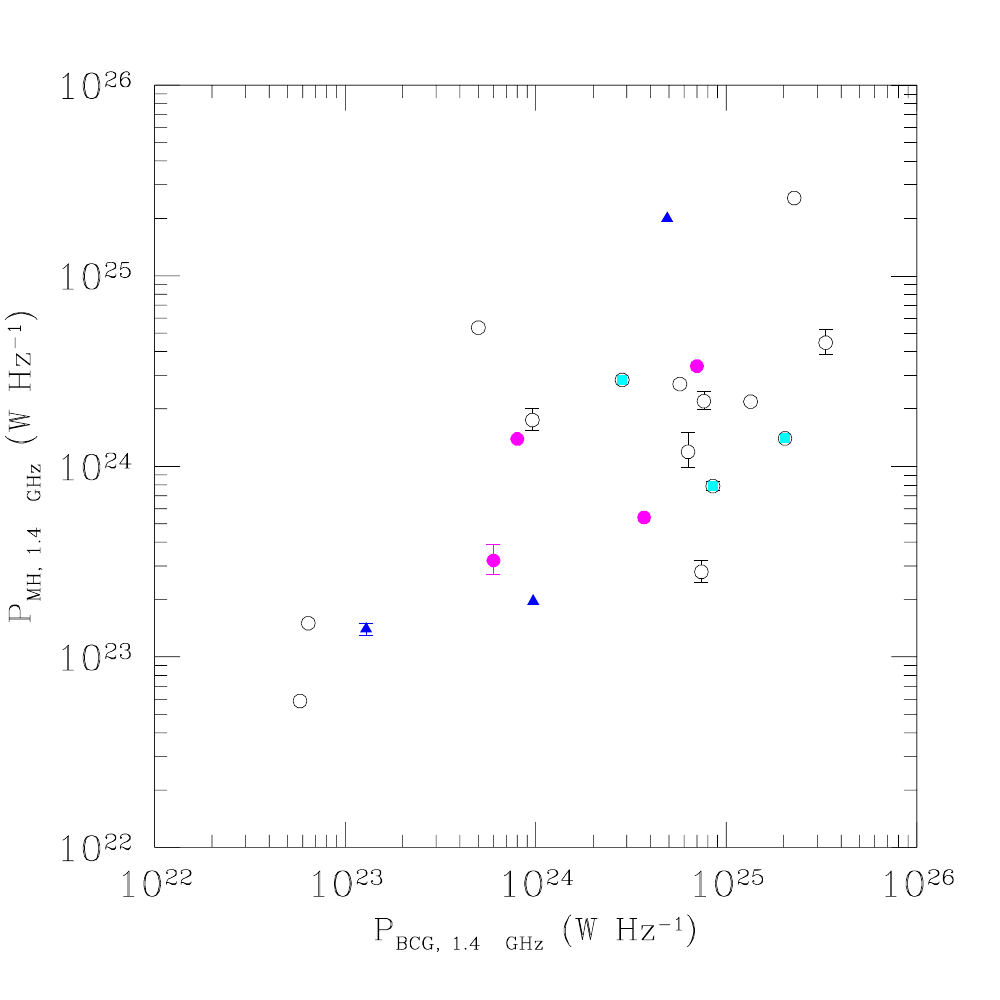} 
\smallskip
\caption{
$P_{\rm MH, \, 1.4 \, GHz}-P_{\rm BCG, \, 1.4 \, GHz}$
diagram for the clusters with previously known minihalos 
(empty black circle), new minihalo detections (magenta circles), 
minihalo candidates (cyan circles) and central extended sources 
whose classification as a minihalo is uncertain (blue triangles). 
Only error bars corresponding to an uncertainty $>10\%$ on 
the radio power are plotted.}
\label{fig:prad1}
\end{figure}

By comparing the radio power of the BCGs and the radio luminosity of the 
surrounding minihalos in 6 clusters, 
\cite{2009A&A...499..371G} noticed that
stronger minihalos tend to occurr in clusters with more powerful 
central radio galaxies, suggesting that the minihalo emission
could be partially related to the activity of the central AGN.
For our larger sample of minihalos and candidates, 
we find a possible weak trend, though with a very large scatter 
(Fig.~\ref{fig:prad1}). A similar trend is also 
visible in the flux-flux plane, indicating that the possible 
relation between minihalo and BCG luminosities may be intrinsic.
We find a Spearman rank correlation coefficient $r_s\sim 0.5$ 
in both planes with a probability of no correlation of few \%.

This result suggests that the central AGN 
activity is not directly powering the minihalo emission, 
although it is one of the plausible sources 
of the seed relativistic electrons for the reacceleration models
\citep{2008A&A...486L..31C}. 
As argued by \cite{2009A&A...499..371G}, a
tight correlation between the minihalo and BCG radio 
properties is not expected, since the radio galaxy likely 
undergoes multiple cycles of activity within the lifetime 
of the minihalo, as supported by the evidence for recurrent 
radio outbursts for a number of cluster and group dominant 
galaxies \citep[e.g.,][Venturi et al. 2013]{2009ApJ...697.1481C,2009ApJ...705..624D,2011ApJ...732...95G,2011ApJ...726...86R,2012ApJ...755..172G}.

%
%
%

%
\begin{figure*}
\centering
\includegraphics[scale=1]{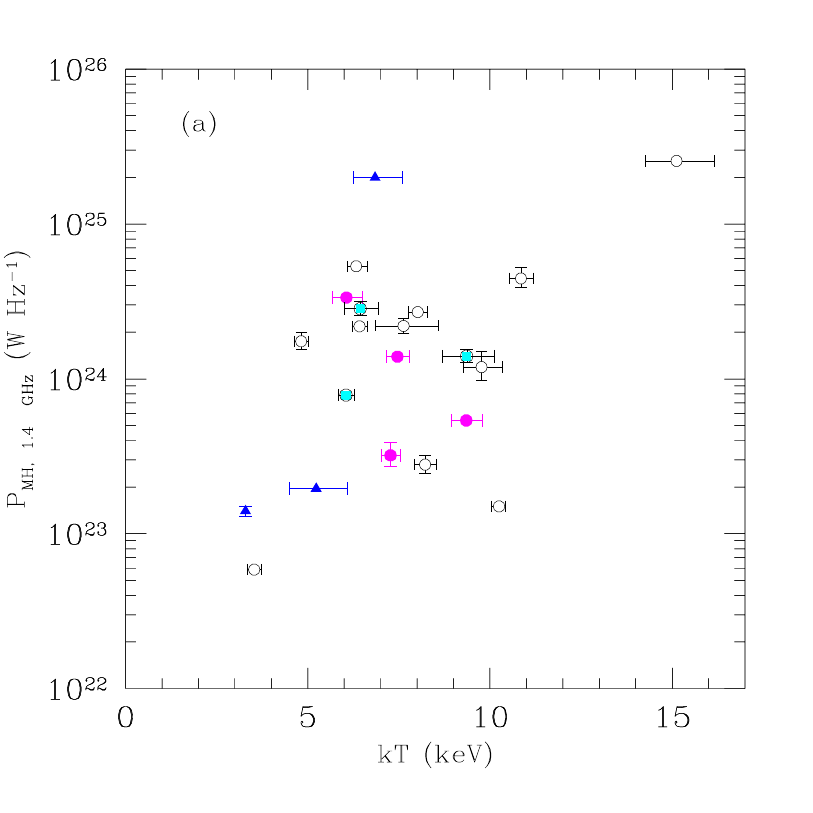}
\includegraphics[scale=1]{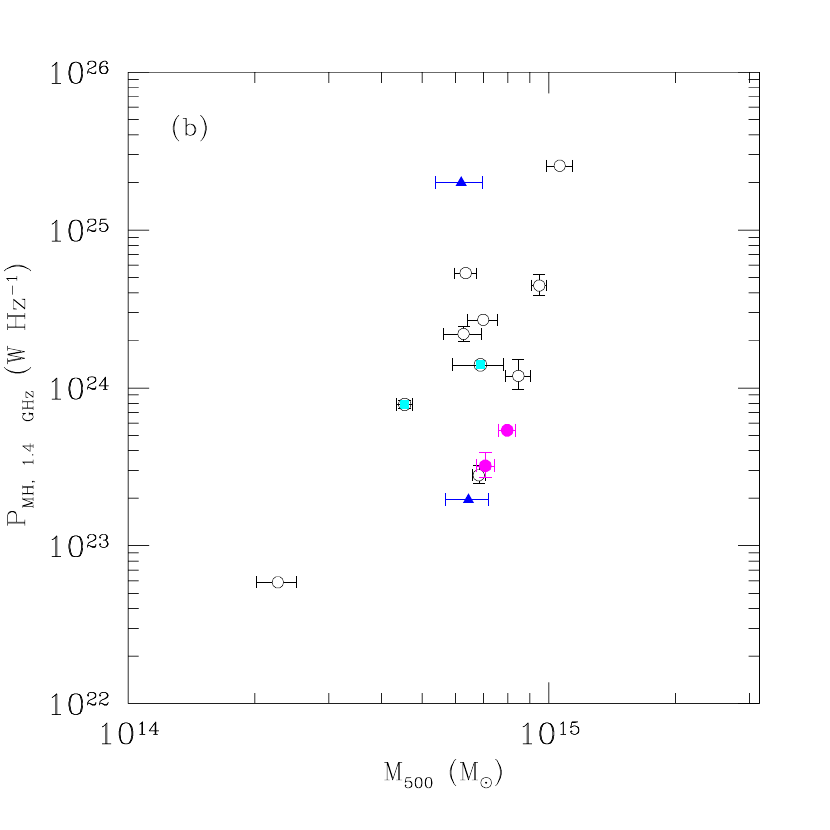}
\smallskip
\caption{
$P_{\rm MH, \,1.4 \, GHz}-kT$ (a) and 
$P_{\rm MH, \,1.4 \, GHz}-M_{\rm 500}$ (b),
diagrams for the clusters with previously known minihalos 
(empty black circle), new minihalo detections (magenta circles), 
minihalo candidates (cyan circles) and central extended sources 
whose classification as a minihalo is uncertain (blue triangles). 
Only error bars corresponding to an uncertainty $>10\%$ on 
the radio power are plotted.}
\label{fig:prad}
\end{figure*}
%
%

\subsection{Comparison between radio minihalos and global properties 
of the cluster hosts}

\cite{2008A&A...486L..31C} found that clusters with higher X-ray luminosity tend 
to possess more powerful minihalos. However, this result was based on only 
six minihalos that were known at the time of their study. A more recent 
investigation was carried out by \cite{2013A&A...557A..99K} for a larger 
number of minihalo clusters (11, of which 5 were from the GRHS cluster sample).
Despite the large scatter, they found indication of a possible radio/X-ray
luminosity correlation, suggesting that an intrinsic relation between the 
thermal and non-thermal cluster properties may exist. 

We can explore this possibility further using our larger sample of 21 
minihalo clusters.
In Fig.~\ref{fig:prad}(a), we plot the minihalo radio power at 1.4 GHz 
versus the cool core-excised cluster temperature (Table 5), 
which can be used as a proxy for the cluster total mass and is strongly
correlated with the X-ray luminosity. Our comparison 
does not indicate a clear scaling between radio power and cluster
temperature, regardless of whether we include candidates or exclude them.
We indeed find $r_s \sim 0.3$ and $P_{\rm no \, corr} \sim 20\%$ 
($r_s \sim 0.1$ and $P_{\rm no \, corr} \sim 70\%$ if we consider only clear minihalo detections).

However, the plot provides us an important piece of information 
on the kind of clusters that possess a minihalo at their center: 
they all tend to have high global temperature, with the majority 
of minihalos found in $T > 5$ keV systems. While proper statistical 
analysis should include non-detections (Giacintucci et al. in prep.), 
this suggests that the hosts of our detected minihalos are massive clusters. 

An alternative way to estimate cluster masses is offered the 
SZ effect \citep[e.g.,][]{2002ARA&A..40..643C}.
Of the 21 minihalo clusters considered here, 14 are in the all-sky cluster catalog 
of validated clusters from the first 15.5 months of {\em Planck} observations 
(Planck Collaboration 2013). In Table 5, we report their total masses within 
$R_{500}$ inferred from the {\em Planck} observations. Fig.~\ref{fig:prad}(b)
shows the distribution of our sub-sample of minihalo clusters that have
{\em Planck} data in the $P_{\rm MH, \,1.4 \, GHz}-M_{\rm 500}$ plane. 
No obvious correlation is visible between the radio luminosity and
cluster mass -- in this case 
we find $r_s \sim 0.3$ and $P_{\rm no \, corr} \sim 10\%$ -- 
in agreement with the lack of a clear 
correlation with the global temperature in panel (a). We note that 
this is in contrast with the {\em giant}\/ radio halos found
in cluster mergers, whose radio luminosity correlates with the cluster mass 
(Cassano et al. 2013 and references therein). 
Again, we find evidence that minihalos are hosted by massive clusters, as all minihalos are 
in $M_{\rm 500} \gax 5 \times 10^{14}$ $M_{\odot}$, except for 
the sligthly less massive system 2A\,0335+096 ($M_{\rm 500} = 2\times 10^{14}$ $M_{\odot}$). 

In a subsequent paper, we will investigate possible correlations with the 
thermodynamical properties of the cool cores rather than the global cluster properties 
(Giacintucci et al. in preparation).

\subsection{Spectral properties of minihalos}\label{sec:alpha}

\begin{figure}
\centering
\includegraphics[scale=0.9]{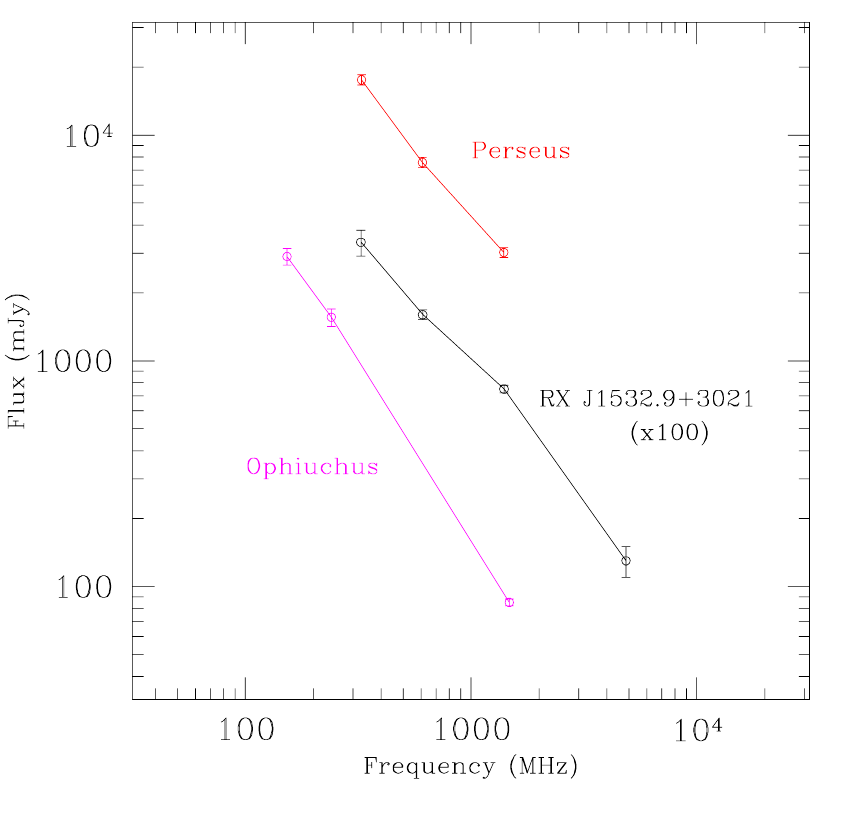}
\smallskip
\caption{Integrated radio spectra of the minihalos in RX\,J1532.9+3021 (black),
Perseus (red, from Sijbring 1993) and Ophiuchus \citep[magenta, from][]{2010A&A...514A..76M}. The total spectral index is $\alpha=1.20\pm0.03$ for RX\,J1532.9+3021,
$\alpha=1.21\pm0.05$ for Perseus and $\alpha=1.56\pm0.04$ for Ophiuchus.
Note that the spectrum of RX\,J1532.9+3021 has been multiplied by 100.}
\label{fig:rxcj1532_sp}
\end{figure}

Until now, our knowledge of the radio spectra of minihalos has been 
limited to only two sources, Perseus and Ophiuchus, 
whose integrated spectra are based on flux density measurements 
at three frequencies. Fig.~\ref{fig:rxcj1532_sp} 
shows these spectra from Sijbring (1993) and \cite{2010A&A...514A..76M}.
In black, we show the spectrum of new minihalo detected
in RX\,J1532.9+3021, based on flux densities at 
four frequencies, reported in \S \ref{sec:1532} (note that 
the spectrum has been multiplied by 100 for display purposes).
The three spectra appear very similar in shape, 
at least in the range of frequencies currently explored. 
They all seem to be well described by a power law with a
steep spectral index $\alpha=1.21\pm0.05$ and 
$\alpha=1.20\pm0.07$ for Perseus and RX\,J1532.9+3021, and
a slighly steeper slope for Ophiuchus, $\alpha=1.56\pm0.04$.
For Ophiuchus, a steepening at the high frequency may be present, 
although the spectral indices below and above the data point at 610 MHz are 
consistent within the errors \citep{2010A&A...514A..76M}. A  
steepening may be also present in the spectrum of RX\,J1532.9+3021, where 
the spectral index changes from $\alpha=1.02\pm0.10$ between 325 MHz and 1.4 GHz 
to $\alpha=1.41\pm0.13$ above 1.4 GHz.

The existing spectral information is not sufficient to
discriminate between the competing models 
for the minihalo formation, i.e., a power-law spectrum 
over the entire radio frequency range expected in pure secondary 
models vs a high-frequency break 
predicted by turbulent reacceleration models. More 
data points and a frequency range wider
than that in Fig.~\ref{fig:rxcj1532_sp} are 
necessary to accurately determine the shape of the
minihalo spectra and confirm the high-frequency
steepening in Ophiuchus and RX\,J1532.9+3021.

%
%
%
%

\subsection{Is particle acceleration needed in small 
size minihalos?}\label{sec:reacc}

As for the Mpc-size, giant radio halos found in many massive, merging
clusters, the few hundred kpc scale of large minihalos (such as Perseus and Ophiuchus) requires 
that the relativistic electrons are continuously injected 
and/or reaccelerated in situ throughout the large emitting volume. 
Indeed, in the strong magnetic fields expected in 
the cool cores, the radio-emitting electrons cannot 
diffuse from the central radio galaxy out to the minihalo radius 
within their radiative lifetime, which is amost two orders of magnitude 
shorter than their diffusion timescale \citep[e.g.,][]{1977ApJ...212....1J}.

The question we pose here is whether diffusion can still account for smaller
minihalos, such as the 
$R_{\rm MH} \sim 50$ kpc emission in A\,2204? The typical diffusion 
velocity of relativistic electrons in clusters is expected to be 
approximately the Alfv\'en velocity \citep{1977ApJ...212....1J}
$v_A = B/(4\pi\rho)^{(1/2)}$, where $B$ is the cluster magnetic field
and $\rho$ is the density of the ICM. Although our knowledge 
of the magnetic field in clusters is still poor, studies of the 
rotation measure of radio galaxies in or behind clusters indicate 
that the magnetic field intensity in cool cores can be as high as 
$5-10$ $\mu$G or even stronger \citep[e.g.,][]{2002MNRAS.334..769T,2004JKAS...37..337C}.
In these strongly magnetized, high-density ($n_{\rm e} \sim 10^{-2}$ cm$^{-3}$)
cluster regions, $v_A \sim 100$ km s$^{-1}$.
At this speed, the time required for the electrons to reach a radius of 
$\sim 50$ kpc from the center is $\sim 5 \times 10^8$ years, which
needs to be compared to their synchrotron radiative cooling time of 
$\sim 10^7$ yr.

Thus, even the small minihalos 
like the one in A\,2204 cannot be produced by electrons leaking 
from the central radio galaxy and diffusing within the cluster 
core volume. However, other physical mechanisms may be able to transport the 
particles from the radio galaxy volume to larger radii faster.
For instance, large-scale sloshing motions 
can potentially advect the electrons along with the gas and redistribute 
them throughout the sloshing region \citep[e.g.,][see also discussions in Gastaldello et al. 2013 and Venturi et al. 2013]{2013ApJ...762...78Z}.
If we assume a sloshing velocity of half the sound speed for a rough estimate,
the electrons could be carried out to a distance of $\sim 30-40$ 
kpc from the center within $10^7$ years, suggesting that this kind of 
transport mechanism may indeed play a role in the less-extended minihalos.

\subsection{ZwCl\,1742 and MACS\,J1931: a radio minihalo, 
fading radio lobes or something in between?}\label{sec:1931}

As shown in \S \ref{sec:1742} and \S \ref{sec:1931}, the radio information 
available for the extended radio sources in the cores of ZwCl\,1742.1+3306 
and MACS\,J1931.8--2634 is not sufficient to unambigoulsy classify the 
central emission as a minihalo; further observations are needed to 
understand their nature.

In MACS\,J1931.8--2634, the diffuse emission is rather elongated 
and seems confined, despite its amorphous appearence. 
Part of the emission is found to be co-spatial with two
possible small, inner X-ray cavities, aligned in the same E-W
direction as the larger scale radio emission. A further complication 
is that the emission at 
150 MHz is significantly more extended than that at higher frequencies;
the size of the source reaches $\sim 600$ kpc in diameter, 
compared to $\sim 200$ kpc 
at 1.4 GHz, and its major axis is slightly tilted with respect to the 
major axis of the high-frequency emission.  Furthermore, 
the total spectrum is found to have an ultra-steep slope ($\alpha \sim 2$). 
As seen in \S \ref{sec:alpha}, minihalos tend to have a spectrum with $\alpha\sim 1.2-1.5$, 
at least in those few cases where spectral information is available.
Ultra-steep spectra are instead common for dying or 
restarted radio galaxies, whose aged emission is no longer 
fed by the nucleus, rapidly fading \citep[e.g.,][Venturi et al. 2013 and references therein]{2009BASI...37...63S,2011A&A...526A.148M,2012ApJ...755..172G}. 
Thus, the ultra-steep spectrum argues 
against a minihalo for an earlier radio outburst of the BCG, which is 
now undergoing a new phase of activity. This is similar to what is seen, 
for instance, at the center of the NGC\,1407 group \citep{2012ApJ...755..172G}.
The different orientation of the low- and high-frequency extended 
structures may also indicate two different (past) radio outbursts that 
occurred at different jet angles. 

On the other hand, the X-ray data indicate that the core of 
MACS\,J1931.8--2634 is sloshing 
in the N-S direction. For a similar sloshing configuration, the simulated 
minihalo maps by \cite{2013ApJ...762...78Z} strikingly resemble 
the radio emission seen at center of MACS\,J1931.8--2634 (see their Figure 11): 
the diffuse emission tends to be distributed in two bright regions 
symmetrical with respect to the cluster center and aligned in a direction 
roughly perpendicular to the axis of sloshing.

The source in ZwCl\,1742.1+3306 could also be classified as a restarted 
radio galaxy. However, no detailed spectral information is available to rule out
the minihalo interpretation. We notice that the source is only $\sim 40$ 
kpc in radius and, as discussed in \S \ref{sec:reacc}, electron reacceleration 
may not be needed to explain the presence of diffuse emission on such a scale. 
If relativistic electrons have been injected by the central AGN during a former 
cycle of activity and accumulated in two radio bubbles, it is possible that  
ongoing sloshing motions have disrupted such bubbles and carried the aged, 
but still radio-emitting, electrons throughout the sloshing region. Thus, 
this object may be an intermediate case between a ``pure'' minihalo emission, arising
from reaccelerated electrons, and an active extended radio galaxy that is 
perturbed, but not disrupted, by large-scale sloshing motions, as 
possibly seen in the WAT at the center of A\,2029 \citep{2004ApJ...616..178C,2013ApJ...773..114P} and in the radio galaxy at the center of A\,3560 (Venturi et al. 2013).

For both clusters, further multi-frequency radio 
observations are needed to discriminate between these interpretations 
by deriving the spectrum of the various contributors to the total
emission and by mapping the distribution of the spectral index.

\section{Summary and conclusions}\label{sec:summ}

We have undertaken a systematic search for new radio minihalos in a large
sample of X-ray luminous clusters with available high-quality {\em VLA}\/ and/or {\em GMRT}\/ 
radio data. In this paper, we present new minihalos and minihalo candidates found in 9
clusters in the course of our re-analysis of the archival {\em VLA}\/
observations. In particular, we found new minihalos in the cores of A\,478, 
RX\,J1532.9+3021, ZwCl\,3146 and A\,2204, and a candidate minihalo, with an 
unusually filamentary morphology, in A\,1795. Diffuse radio emission is 
also found in the cores of MACS\,J0159.8--0849 and MACS\,J0329.6--0214. 
The existing data for the latter 2 clusters do not allow us to rule out 
the possibility of this emission being part of their central radio galaxies, 
therefore we list them among the minihalo candidates. Other new minihalo 
candidates include ZwCl\,1742.1+3306 and MACS\,J1931.8--2634; further radio 
observations are required to determine if these are restarted radio galaxies 
or minihalos.

The radio luminosities of our minihalos and candidates range between
$10^{23}$ W Hz$^{-1}$ and $10^{25}$ W Hz$^{-1}$ at 1.4 GHz, as commonly
found for this type of radio sources. Their sizes ($R_{\rm MH}=40\div160$ kpc) are 
somewhat smaller than those of the previously known minihalos, which, in some cases, 
extend to $\sim 300$ kpc from the center.

Combining our new detections with the previously known minihalos results in
a total sample of 21 minihalos. We briefly compared radio properties of
these clusters with their global X-ray temperatures and total masses
estimated from {\em Planck} observations. While proper statistical analysis
should include non-detections (Giacintucci et al.\ in prep.), we found
that clusters hosting minihalos tend to be hot ($T\gax 5$ keV) and
massive systems. Beyond that, we did not see any clear correlation between
the minihalo radio luminosity and the cluster global temperature or total
mass, in contrast with the behavior of the {\em giant}\/ radio halos in
merging clusters, whose radio power correlates with the cluster mass (e.g.,
Cassano et al.\ 2013). In a subsequent paper, we will investigate 
possible correlations with thermodynamical properties of the cool cores 
rather than the global cluster properties.

{\em Chandra}\/ X-ray images indicate the presence of cold
fronts and gas sloshing in the cores of most of our clusters. The minihalos
are often (though not always) contained within the regions bounded by the
sloshing cold fronts, as previously observed in other minihalo clusters
(e.g., Mazzotta \& Giacintucci 2008; ZuHone et al.\ 2013, Giacintucci et
al.\ in prep.). This supports the hypothesis that radio emitting electrons
are reaccelerated by turbulence generated by sloshing (ZuHone et al.\ 2013).
We also argue that simple advection of radio-emitting electrons by the
sloshing gas (without the need for reacceleration) may play a role in the
formation of the less-extended minihalos.
\\
\\
{\it Acknowledgements.}

The authors thank the anonymous referee, whose comments and suggestions
improved the paper. SG acknowledges the support of NASA through Einstein Postdoctoral Fellowship PF0-110071 awarded by the Chandra X-ray Center (CXC), which
is operated by SAO. Basic research in radio astronomy at the
Naval Research Laboratory is supported by 6.1 base funding.
The National Radio Astronomy Observatory is a facility of the National Science 
Foundation operated under cooperative agreement by Associated Universities, Inc.
This research work has used the TIFR GMRT Sky Survey (http://tgss.ncra.tifr.res.in) data products.

{}

\appendix

\section{The radio source {\em VLA}--J1532+3018 in the field of 
RXJ\,1532.9+3021\label{sec:appendix}}

\begin{figure}
\centering
\includegraphics[width=10cm]{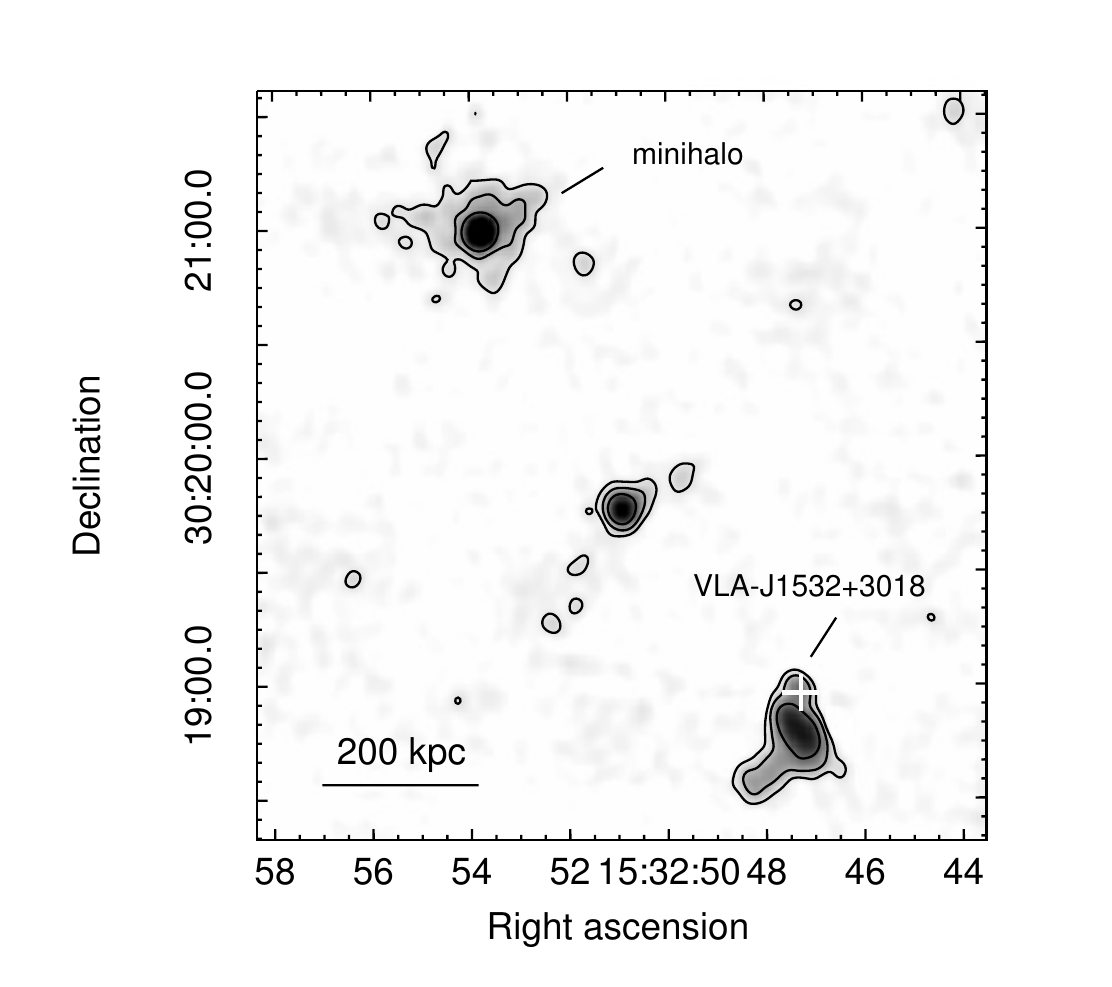}
\smallskip
\caption{{\em VLA} B-array image at 1.4 GHz of the field of RX\,J1532.9+3021, 
hosting a radio minihalo at its center (see also Fig.~2). A radio source 
with complex morphology ({\em VLA}-J1532+3018) is located $\sim 2.6^{\prime}$ south-east 
of the RX\,J1532.9+3021 center. The source is associated with the optically-selected 
cluster GMBCG\,J233.19725+30.31626, whose center is marked as a white cross. The restoring 
beam is $4.2^{\prime\prime}\times3.7^{\prime\prime}$, 
in p.a. $-11^{\circ}$ and contours are 0.045, 0.18 and 0.72 mJy beam$^{-1}$.}
\label{fig:field}
\end{figure}

\begin{figure*}
\centering
\includegraphics[width=7cm]{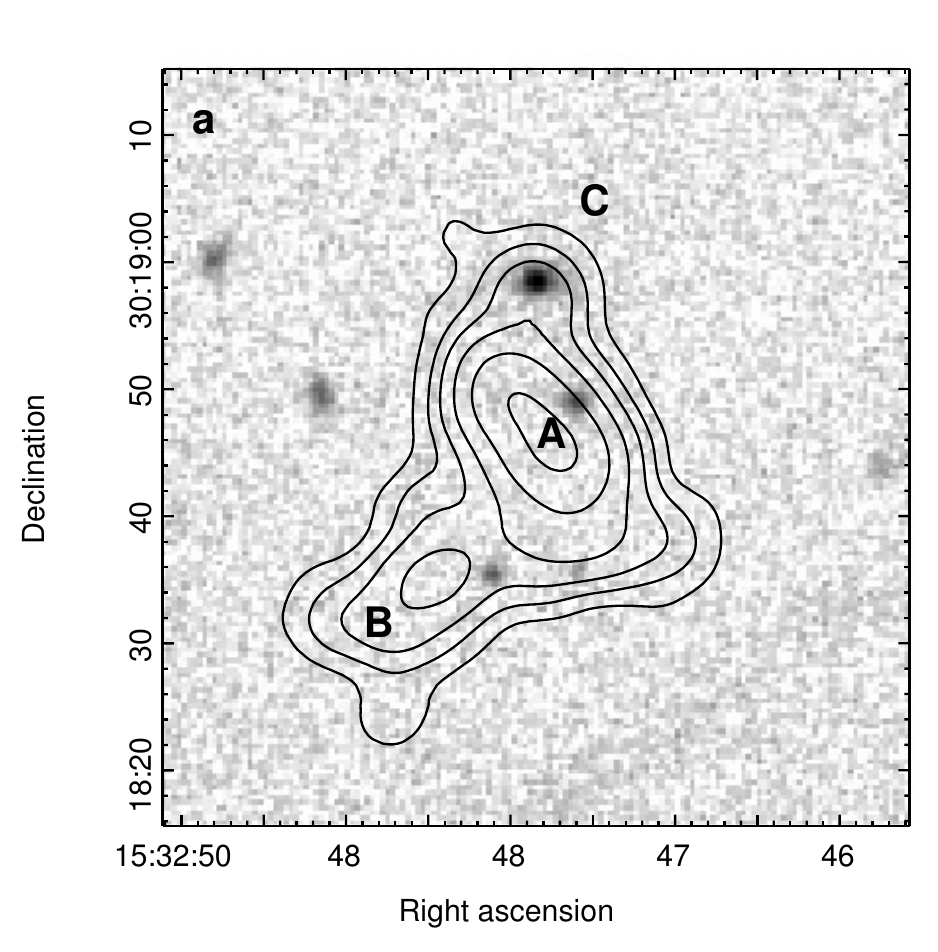}
\hspace{0.5cm}
\includegraphics[width=7cm]{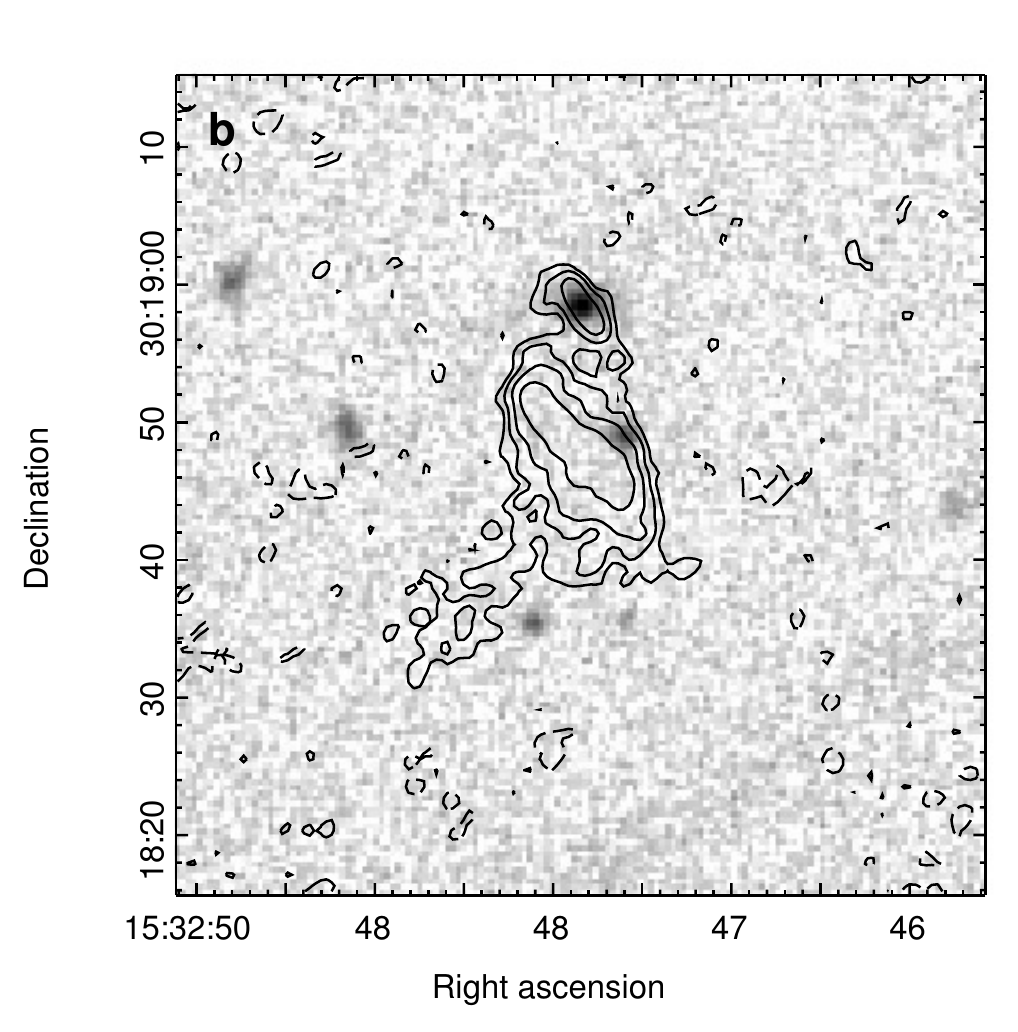}
\smallskip
\caption{(a) {\em GMRT} contour image of {\em VLA}-J1532+3018 at 610 MHz, 
overlaid on the optical SDSS image. The restoring beam is
  $6.0^{\prime\prime}\times5.0^{\prime\prime}$, in p.a. $85^{\circ}$
and contours are 0.12, 0.24, 0.48, 0.96, .... mJy beam$^{-1}$. No
levels at $-0.12$ mJy beam$^{-1}$ are
present in the portion of the image shown. Labels indicate the
different components of the source. (b) {\em VLA} 1.4 GHz contours from  
the combined A+B-array image of the same region as panel (a). 
The restoring beam is 
$1.6^{\prime\prime}\times1.4^{\prime\prime}$, in p.a. $-32^{\circ}$,
and contours are $-0.04$ (dashed), 0.04, 0.08, 0.16, 0.32, ...
mJy beam$^{-1}$.}
\label{fig:s_source}
\end{figure*}

An extended radio source with complex morphology is located $\sim
2.6^{\prime}$ south-west of the center of the minihalo-cluster RXJ\,1532.9+3021 
(Fig.~\ref{fig:field}).  Hereinafter, we refer to this source as {\em VLA}-J1532+3018.
The {\em GMRT} image at 610 MHz \citep{2008A&A...484..327V} 
and {\em VLA} 1.4 GHz image from
the combined A+B array (Table 2) are shown in
Fig.~\ref{fig:s_source}. Both images are overlaid as contours on the optical red
band image from the SDSS. At 610 MHz (panel (a)), three main components
can be seen: a bright central component, labelled A, that is  
mostly elongated in the NE-SW axis; an extended spur-like feature toward SE (B);  
a peak of emission to the north (C), coincident with a bright optical galaxy.

At higher resolution (panel (b)), C is resolved into an extended source with a 
south-west to north-east major axis. The source is clearly associated with the galaxy
SDSS J153247.33+301858 with $z_{\rm phot} = 0.35\pm0.03$.
The central component A appears as a double-lobe radio galaxy, 
though no clear jets or a compact central core are visible.
The emission in B has very low surface brightness and is apparently 
connected to A in the form of a tail. The optical galaxy SDSS J153247.09+301849 
($z_{\rm phot} = 0.36\pm0.03$) is located within the radio contours, on the western side 
of A, but its association with the radio source is uncertain. 
A and C are also detected, with similar morphology, in the images at 4.9 GHz 
and 8.5 GHz (not shown here; see Table 2 for details), while, for region B, 
only the emission coincident with the 610 MHz peak is visible at 4.9 GHz, and
no emission is detected at 8.5 GHz.

We interpret {\em VLA}-J1532+3018 as the blend of the radio galaxy B and
a possible second radio galaxy (A), located at approximately the same redshift. 
This latter may be a tailed radio source with both jets bent to 
form a tail toward SE (C). Assuming $z=0.36$, the lenght of the tail is $\sim 100$ kpc.
A summary of the properties of {\em VLA}-J1532+3018 is provided in Table 7.

Tailed radio galaxies are typically members of groups or clusters of galaxies. 
We searched therefore the literature for indication of a group/cluster at the 
position of {\em VLA}-J1532+3018 and found an optically-selected galaxy 
cluster GMBCG\,J233.19725+30.31626 at $z_{\rm phot} = 0.358\pm0.034$, 
with $\sim 10$ possible member galaxies \citep{2010ApJS..191..254H} 
and a low-significance X-ray enhancement in the {\em Chandra} image.
The cluster is at similar distance as RX\,J1532.9+3021 ($z=0.36$; Table 1) 
and their separation in the plane of the sky is approximately 800 kpc.

\begin{table}
\caption[]{Properties of VLA-J1532+3018}
\begin{center}
\begin{tabular}{cccccccccc}
\hline\noalign{\smallskip}
 Component  & optical id & $z_{\rm phot}$ & $S_{\rm 610 \, MHz}$ & $S_{\rm 1.4 \, GHz}$ & $S_{\rm 4.9
  \, GHz}$ & $S_{\rm 8.5  \, GHz}$ & $\alpha$ & $P_{\rm 1.4 \, GHz}$ &
Size \\
               &  (SDSS)    & &  (mJy) & (mJy) & (mJy) & (mJy) & & ($10^{24}$ W Hz$^{-1}$) & (kpc$\times$kpc)\\
\noalign{\smallskip}
\hline\noalign{\smallskip}
A+B & \phantom{0}J153247.09+301849 $^a$ & $0.36\pm0.03$ & $25.0\pm1.3$ &

$13.6\pm0.7$ & $4.7\pm0.2$ & $2.1\pm0.1$ & $0.94\pm0.03$ & $6.0\pm0.3^b$ & $70\times240^b$ \\

C & J153247.33+301858 & $0.35\pm0.03$& $2.0\pm0.1$  &
$1.50\pm0.08$ & $1.00\pm0.05$ &
 $0.71\pm0.04$  & $0.39\pm0.03$ &$0.62\pm0.03$ & $15\times30$ \\
\hline{\smallskip}
\end{tabular}
\end{center}
\label{tab:src}
Notes to Table \ref{tab:src} -- $(a)$: uncertain identification. $b$:
assuming $z_{\rm phot} = 0.36$. 
\end{table}

%
%
%

%
%

\end{document}